\documentclass[
 reprint,=
 amsmath,amssymb,
 aps,superscriptaddress]{revtex4-2}
\usepackage{xcolor}
\usepackage{graphicx}
\usepackage{dcolumn}
\usepackage{bm}

\newcommand{\be}{\begin{equation}}
\newcommand{\ee}{\end{equation}}

\usepackage{multirow}
\usepackage{color}
\usepackage{xcolor}
\usepackage{times}
\usepackage{amsmath,bm,amsfonts}
\usepackage{dcolumn}
\usepackage{latexsym}
\usepackage{hhline}
\usepackage{braket}
\usepackage{bbold}

\usepackage{adjustbox}
\usepackage{dsfont}
\usepackage{xcolor}
\usepackage{amsmath,amsfonts, amssymb, amsthm, dsfont}
\usepackage{yfonts}
\usepackage{bm}
\usepackage{mathrsfs}
\usepackage{graphicx}
\usepackage{verbatim}
\usepackage{hyperref}
\usepackage{wasysym}
\usepackage{tikz}
	\usetikzlibrary{calc}

\renewcommand{\vec}[1]{{\mathbf #1}}

\renewcommand{\vr}{{\vec{r}}}

\newcommand{\comments}[1]{}

\def\Z{\mathbb{Z}}

\begin{document}

\title{Topological order and Fractons from Gauging Exponential Symmetries}

\author{Guilherme Delfino}
\affiliation{Department of Physics, Boston University, MA, 02215, USA}
\author{Claudio Chamon}
\affiliation{Department of Physics, Boston University, MA, 02215, USA}
\author{Yizhi You}
\affiliation{Department of Physics, Northeastern University, MA, 02115, USA}

\date{\today}
\begin{abstract}
We broaden the scope of quantum field theory by introducing a general class of discrete gauge theories that realize either topological order or fracton behavior across dimensions. We start from translation-invariant systems endowed with unconventional charge-conservation laws, which we term \textit{exponential polynomial symmetries}. Gauging these symmetries yields $\mathbb{Z}_N$ gauge theories in 2D that exhibit topological order whose quasiparticles have constrained mobility and whose ground-state degeneracy shows ultraviolet (UV) dependence. These features are reminiscent of spatial symmetry–enriched topological order, wherein quasiparticle excitations transform nontrivially under lattice translations. We further propose a Chern–Simons variant that produces non-CSS stabilizer codes and outline a framework for exponentially symmetric subsystem SPT phases. Finally, we extend this gauging procedure to 3D, obtaining new variants of fracton topological order.
\end{abstract}

\maketitle

\tableofcontents
\section{Introduction}

After several decades of progress, the concept of topological order \cite{Wen04,W9505,SB160103742} has become a widely accepted framework for describing a broad range of quantum states of matter. Quantum stabilizer codes have played a key role in understanding topological order through exactly solvable Hamiltonians. Kitaev's toric code \cite{kitaev03} and Wen's plaquette model \cite{wen03} are prime examples: they capture the essential physics of discrete $\mathbb{Z}_N$ gauge theories, feature topological quasiparticles, and exhibit ground-state degeneracy originating from an exactly solvable many-body spectrum. These models have also deepened our understanding of topological entanglement entropy \cite{levin-wen,kitaev-preskill}, braiding statistics, and holonomies.

In this work, we extend the scope of topological field theories by introducing a broader class of discrete gauge theories governed by unconventional charge-conservation laws \cite{P160405329,SPP180700827,PY200101722,barkeshli,MHC180210108,oh22a,pace2022position,oh22b,seiberg22a,Ebisu2023}:
\begin{eqnarray}\label{exponential}
G[f,g] = \sum_r g_r \, a^{f_r} \, q_r
\end{eqnarray}
which we refer to as \textit{exponential polynomial charges}. Here, $G[f,g]$ generates continuum spatially modulated symmetries \cite{sala2022dynamics,Lehmann2023,hu2023spontaneous,Lian2023}; $q_r$ is the charge density at lattice site $r$; $f_r$ and $g_r$ are integer-valued functions with polynomial dependence on $r=(x,y)$; and $a$ is an integer parameter defined mod $N$.
Gauging these \textit{exponential polynomial symmetries} in 2D yields $\mathbb{Z}_N$ gauge theories, and the resulting 2D topologically ordered states exhibit several distinctive features\cite{williamson2019fractonic,Watanabe2022,oh2022effective,oh2023aspects,pace2022position,kim2025noninvertible,pace2025spacetime,ebisu2024foliated,ebisu2024foliated,bulmash2025defect,seo2025noninvertible,delfino2024anyon}: quasiparticles with restricted dynamics and a torus ground-state degeneracy that exhibits ultraviolet (UV) sensitivity due to the non-commutativity between topological excitation operators and lattice translations\cite{SS200310466,GLS210800020,pace2022position,seiberg22a,delfino22,williamson19,radzihovsky,slagle2017quantum,you2020fractonic,haah2011local,prem2019cage,shirley2020twisted,oh22a,oh22b}. We will show that these $\mathbb{Z}_N$ gauge theories can be interpreted as spatial symmetry–enriched topological orders in two dimensions.

Following this spirit, we extend our theory to higher dimensions by implementing an analogous 3D construction that produces a new class of fracton topological orders in which subsystem symmetries intertwine with exponential symmetries\cite{ma2018fracton,bulmash2018higgs,Lake2021,oh22a,seiberg22a,pace2022position,delfino22,li2020fracton,huang2023chern,Vijay2016-dr,2018arXiv180302369Y,devakul2018strong,SS200310466,aasen2020topological,seo2025noninvertible,li2020fracton,kim2025unveiling}.
In general, the gauge theory obtained by gauging the exponential symmetry exhibits the following features:
\begin{enumerate}
\item Certain deconfined charge or flux excitations have restricted motions.
\item The ground state exhibits long-range entanglement; its degeneracy oscillates with system size.
\item Anyon excitations display nontrivial, path-dependent braiding statistics.
\end{enumerate}

To scrutinize the topological and fracton orders emerging from gauging exponential polynomial symmetries, the paper is organized as follows. In Sec.~\ref{sec:one}, we construct generalized $2$D gauge theories with exponential charge symmetries and show how they can arise from gauging bosonic lattice models, relating them to the models of Ref.~\cite{Watanabe2022}. In Sec.~\ref{sec:two}, we analyze the low-energy physics of a concrete $\mathbb{Z}_N$ gauge theory from exponential symmetries, including its long-range entanglement, constrained excitations, and oscillatory but finite ground-state degeneracy (GSD). In Sec.~\ref{beyond}, we discuss extensions to Chern–Simons-like theories and subsystem SPT phases in two dimensions. Finally, we generalize the framework to three dimensions by incorporating exponential symmetry into the X-cube model, yielding a broad class of fracton gauge theories.

\section{2D Discrete Gauge Theory from gauging Exponential Symmetry }\label{sec:one}

\subsection{Gauging exponential charge symmetry}

As a warm-up, we formulate a generalized electromagnetic framework that incorporates exponential charge conservation and show that the resulting gauge theory exactly realizes the Fuji–Cheng–Watanabe stabilizer code of Ref.~\cite{Watanabe2022}.
To analyze the charge pattern in an electromagnetic theory on a 2D square lattice, we define a two-component `electric field', denoted by $E_{1,\ell}$ and $E_{2,\ell}$, which reside on the $x$- and $y$-links of the lattice. These fields obey a special Gauss law specified by two integer parameters, $a_1$ and $a_2$. The charge $q_r$ lives at the vertex $r$ of the square lattice:
\begin{align}
q_r=
E_{1,r-\frac{e_x}{2}}-a_1 E_{1,r+\frac{e_x}{2}}+E_{2,r-\frac{e_y}{2}}-a_2 E_{2,r+\frac{e_y}{2}}
\label{exp:gua}
\end{align} 
At this point, we assume that the system carries a $U(1)$ charge for generality, though we will later focus on the case of discrete charges.
In the above, $e_x$ and $e_y$ denote the lattice unit vectors along the $x$ and $y$ directions, and $r$ labels the position of the vertex where charge lives.
Eq.~\eqref{exp:gua} can be interpreted as a generalized Gauss law incorporating exponential charge conservation, subject to suitable boundary conditions. While the total charge density $\sum_r q_r$ is not conserved, the theory preserves an \textit{exponential charge conservation} \cite{sala2022dynamics,Lehmann2023, Watanabe2022, Lian2023}, where
\begin{align}
G = \sum_r a_1^x a_2^y  q_r
\label{con}
\end{align}
is conserved. Suppose we consider the isotropic case with $a_1=a_2$, then Eq.~\eqref{con} reduces to a special case of the exponential polynomial generators $G[f,g]$ in Eq.~\eqref{exponential} with $g_r=1$ and $f_r = x+y$.

This system can also be viewed as a scalar charge theory in which each particle at position $(x,y)$ carries a site-dependent charge $a_1^x a_2^yq_r$. Consequently, the charge sectors transform nontrivially under translations:
\begin{align}
T_x G T^{-1}_x=a_1 G, \quad T_y G T^{-1}_y=a_2 G.
\end{align}
In this work, we will be primarily interested in discrete charge systems where $G$ is defined modulo $N$ by Higgsing the $U(1)$ group down to $\mathbb{Z}_N$. We will revisit this point in Sec.~\ref{sec:discrete}.

\begin{figure}[h!]
  \centering
      \includegraphics[scale=0.45]{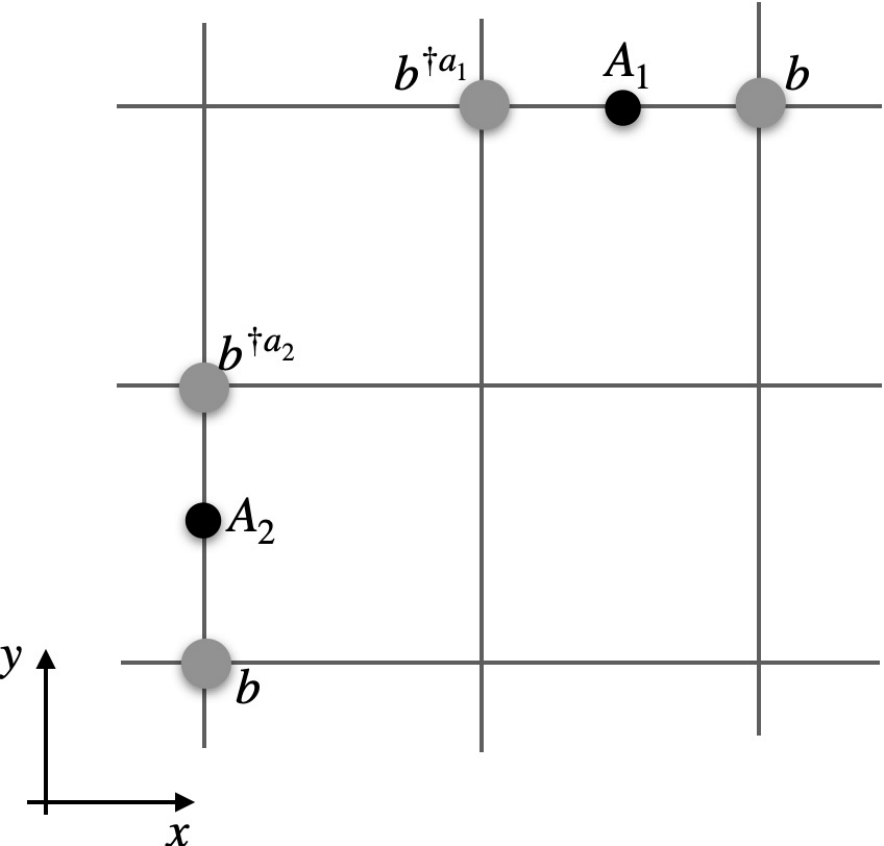}
  \caption{The charged bosonic fields $b$ live at the vertices of the square lattice, while the gauge fields $A_1$ and $A_2$ are defined at the horizontal and vertical edges.} 
\label{bosons_FCW}
\end{figure}

One possible way to describe the charge dynamics on the lattice, subject to the special conservation law in Eq.~\eqref{con}, is via a boson model on a square lattice:
\begin{align}
b_r^{a_1} b_{r+e_x}^{\dagger}+b_r^{a_2} b_{r+e_y}^{\dagger}+\text{h.c.}
\label{co}
\end{align}
Here, $b^{\dagger}_r$ and $b_r$ are boson creation and annihilation operators at site $r$ in the second-quantization formalism. This charge dynamics creates $a_1$ ($a_2$) units of charge at site $r$ while annihilating a unit charge at the neighboring site $r+e_x$ ($r+e_y$). Importantly, the hopping term in Eq.~\eqref{co} does not preserve global charge conservation, but it does respect the exponential charge conservation law in Eq.~\eqref{con}.
To elucidate the gauge structure associated with the exponential symmetry in Eq.~\eqref{con}, we couple the charged boson field to a gauge potential $A_{1,\ell}$ and $A_{2,\ell}$:
\begin{align}
b_r^{a_1} e^{iA_{1,r+\frac{e_x}{2}}} b_{r+e_x}^{\dagger}+b_r^{a_2} e^{iA_{2,r+\frac{e_y}{2}}} b_{r+e_y}^{\dagger}+\text{h.c.}.
\end{align}
The gauge potentials, which are conjugate to the electric fields, reside on the $x$- and $y$-links of the square lattice, as illustrated in Fig.\ref{bosons_FCW}. They transform under gauge transformations of the form
\begin{align}
&A_1 \rightarrow A_1 - f_{r+e_x} + a_1 f_r,\nonumber\
&A_2 \rightarrow A_2 - f_{r+e_y} + a_2 f_r,
\end{align}
which explicitly depend on the parameters $a_1$ and $a_2$.

\begin{figure}[h!]
  \centering
      \includegraphics[scale=0.4]{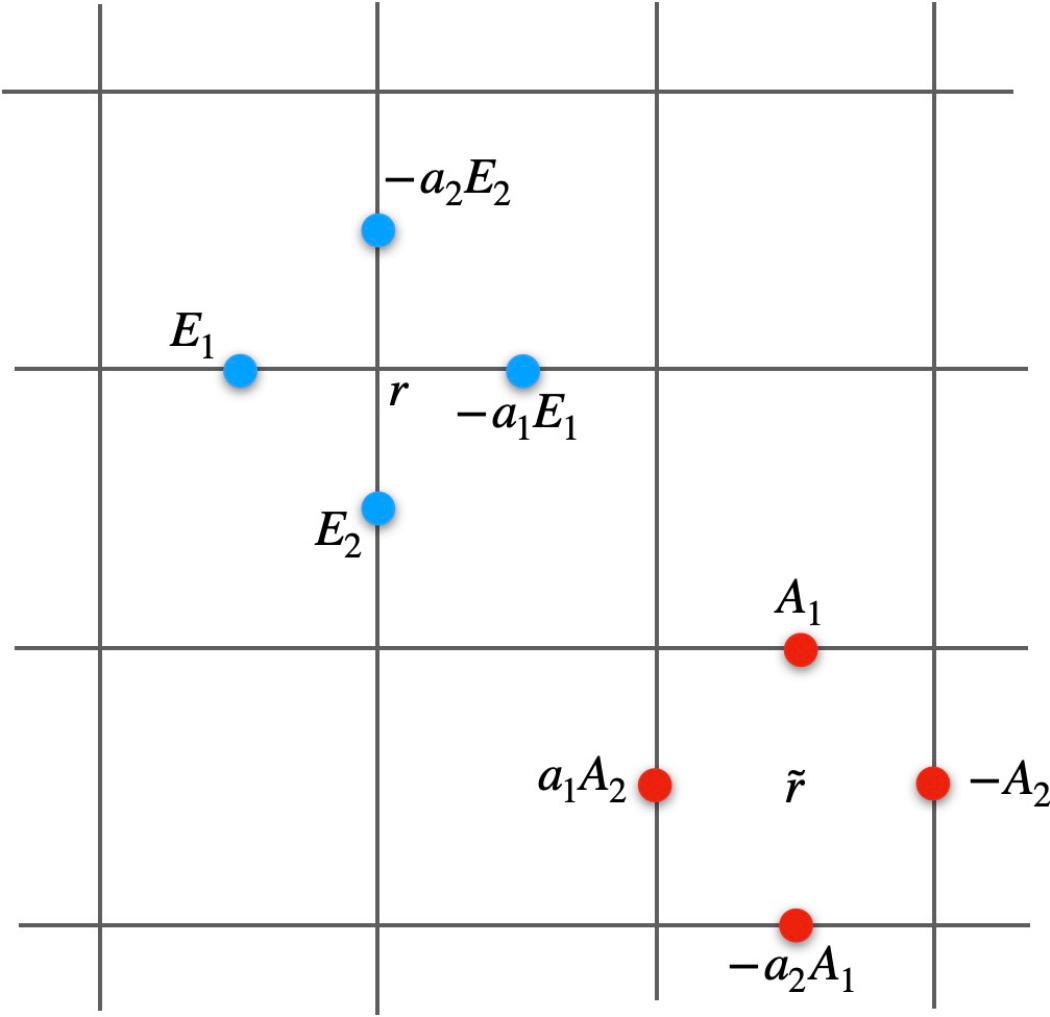}
  \caption{Charge conservation law at $r$ and magnetic flux operators at $\tilde r$ defined on the square lattice.} 
\label{charge_flux_FCW}
\end{figure}

We hereby define the leading order, gauge-invariant, magnetic flux operator,
\begin{eqnarray} 
B_{\tilde{r}}
 = A_{1,\tilde{r}+\frac{e_y}{2}}-a_2 A_{1,\tilde{r}-\frac{e_y}{2}} - A_{2,\tilde{r}+\frac{e_x}{2}}+a_1 A_{2,\tilde{r}-\frac{e_x}{2}}
\label{exp:flux}
\end{eqnarray}
which lives at the center of the plaquette on the dual lattice $\tilde{r}$ illustrated as Fig.~\ref{charge_flux_FCW}. Here $\tilde{r}$ denotes the position of the central plaquette.
To simplify the notation, we introduce two operators $D_i$ and $\Tilde{D}_i$ that can be viewed as generalized differential polynomials operators defined on the original and dual square lattice, respectively:
\begin{eqnarray}
&D_1f_r=-a_1 f_{r+\frac{e_x}{2}}+ f_{r-\frac{e_x}{2}}\nonumber\\
&D_2f_r=- a_2f_{r+\frac{e_y}{2}}+ f_{r-\frac{e_y}{2}}\nonumber\\
&\Tilde{D}_1f_{\tilde r}=- f_{\tilde{r}+\frac{e_x}{2}}+a_1 f_{\tilde{r}-\frac{e_x}{2}}\nonumber\\
&\Tilde{D}_2f_{\tilde r}=- f_{\tilde{r}+\frac{e_y}{2}}+a_2 f_{\tilde{r}-\frac{e_y}{2}}
\end{eqnarray} 
Using this notation, we can express the Gauss law and magnetic flux operators in the compact form:
\begin{eqnarray} 
 D_1 E_1+D_2 E_2&=& q,\nonumber\\
 \Tilde{D}_1 A_2-\Tilde{D}_2 A_1&=&B.
\end{eqnarray}

\subsection{Discretization to $\mathbb{Z}_N$ charge}\label{sec:discrete}

\begin{figure}[h!]
  \centering
      \includegraphics[scale=0.4]{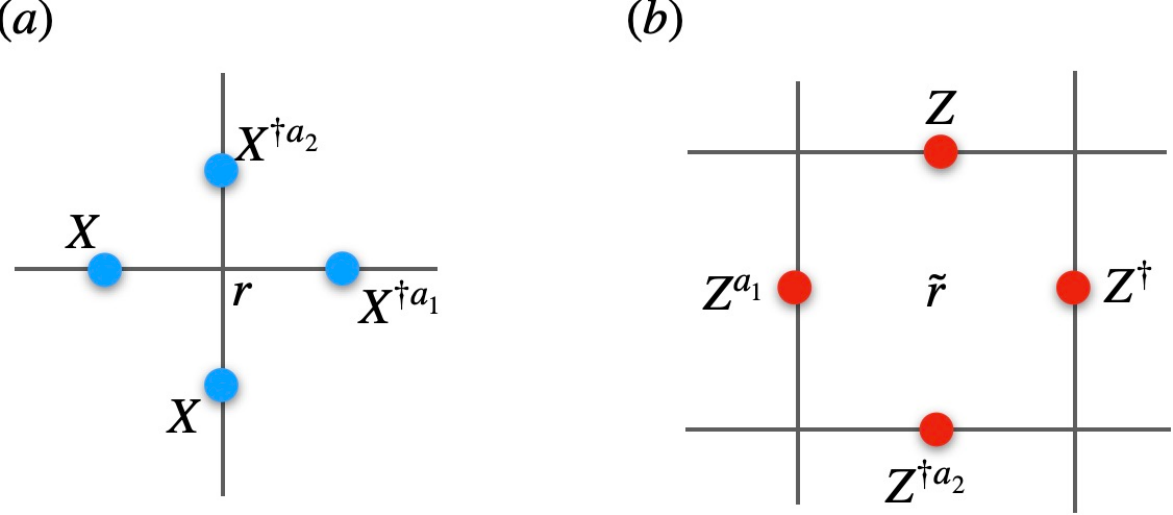}
  \caption{ Operators for the $\mathbb{Z}_N$ (a) charge $Q_r$ and (b) flux $B_{\tilde r}$ at sites $r$ and $\tilde r$ at original and dual lattices, respectively.} 
\label{zn_FCW}
\end{figure}

Our next step is to establish a connection between our theory and the Fuji–Cheng–Watanabe stabilizer code in Ref.~\cite{Watanabe2022}. To discretize our gauge theory, we add a $\cos(Nq_r)$ term to the Hamiltonian, which energetically favors $\mathbb{Z}_N$ charges in the low-energy sector. Consequently, the corresponding electromagnetic fields can be represented using the $\mathbb{Z}_N$ Pauli operators $\omega^E = X$ and $e^{iA} = Z$, with $\omega = e^{2\pi i/N}$. The canonical quantization between the vector potential and the electric field components reproduces the $\mathbb{Z}_N$ Pauli algebra $XZ = \omega  ZX$ for operators on the same edge. Notably, the resulting gauged theory is equivalent to the stabilizer code in Ref.\cite{Watanabe2022}, as
 \begin{align} \label{watanabe}
H=-\sum_r Q_r -\sum_{\tilde r} B_{\tilde r}+\text{h.c.}
\end{align}
where
\begin{eqnarray}\label{chargess}
Q_r &= &X^{-a_1}_{r+\frac{e_x}{2}}X_{r-\frac{e_x}{2}}X^{-a_2}_{r+\frac{e_y}{2}}X_{r-\frac{e_y}{2}}\nonumber\\
B_{\tilde r}&=&Z^{-1}_{\tilde{r}+\frac{e_x}{2}}Z^{a_1}_{\tilde{r}-\frac{e_x}{2}}Z_{\tilde{r}+\frac{e_y}{2}}Z^{-a_2 }_{\tilde{r}-\frac{e_x}{2}}
\end{eqnarray} 
as shown in Fig. \ref{zn_FCW}.
The ground states of the above Hamiltonian are projected onto the vanishing charge and flux sector by the $Q_r$ and $B_r$ operators. 
As discussed in Ref. \cite{Watanabe2022}, the model realizes either topologically ordered phases or trivial phases depending on the parameters $a_1,a_2$ in the model.

First, we comment on the case where $a_i \equiv 0 \ \text{mod} \ \mathrm{rad}(N)$, where $\mathrm{rad}(N)$  denotes radical of an integer N. In this case, the ground-state manifold corresponds to a trivial phase that can be disentangled using finite-depth local unitary circuits, as noted in Ref.~\cite{Watanabe2022}. From the perspective of exponential charge symmetry, this can be understood by recalling that when gauging the matter field in Eq.~\eqref{co}, special care is required if the charge is a discrete $\mathbb{Z}_N$ field.
If $a_i \not\equiv 0 \ \text{mod} \ \mathrm{rad}(N)$, the exponential symmetry operator $G$ in Eq.~\eqref{con} remains a genuine global symmetry, and the gauging procedure can be applied as proposed. However, when $a_i = 0 \ \text{mod} \ \mathrm{rad}(N)$, there exists an integer $m$ such that $a_i^{m} \equiv 0 \ \text{mod} \ N$. In this situation, the exponential symmetry acts nontrivially only on a finite number of degrees of freedom for sites with $|r| < m$, and after coarse-graining, $G$ effectively becomes a local symmetry in the thermodynamic limit. Since a local symmetry cannot be gauged further, no additional gauge structure arises. This explains why the stabilizer code in Ref.~\cite{Watanabe2022} yields a trivial phase when $a_i = \mathrm{rad}(N)$.

For simplicity in the following discussion, and without loss of generality, we \textit{set $a_1 = a_2 \equiv a$} and fix the system size to $L$. After discretizing the charge to $\mathbb{Z}_N$, the exponential symmetry generator in Eq.~\eqref{con} takes the form:
\begin{align}
    G_N=\prod_{r} Q_r^{a^{x+y}}
\end{align}
Under periodic boundary conditions, we require that after translating by $L$ sites (either in the $x$ or $y$ direction), $G_N$ remains well-defined, meaning $a^L = a^0$ mod $N$. When $a^L - 1 = 0$ mod $N$, the exponential symmetry is well-defined on a closed manifold, and the resulting theory is reminiscent of a $\mathbb{Z}_N$ gauge theory. In contrast, when $a^L - 1 \neq 0$ mod $N$, the $\mathbb{Z}_N$ exponential symmetry is effectively reduced to a $\mathbb{Z}_k$ symmetry,
\begin{align}
    G_k=\prod_{r} Q_r^{\frac{N}{k}(a^{x+y})}
\end{align}
with $k=\gcd(a^L-1,N)$ in order to be compatible with the periodic boundary conditions. In the previous expression, $\gcd(A,B)$ stands for the greatest common divisor between the two positive integer numbers $A$ and $B$. The resulting gauge theory should be akin to a $\mathbb{Z}_k$ gauge theory, explaining the size dependence of the ground state degeneracy observed in Ref.~\cite{Watanabe2022}. Here we mention that the result $k=\gcd(a^L-1,N)$ holds only for $a$ and $N$ coprime. We discuss the more general case in Appendix \ref{global_constraints} in the context of the model proposed in Section \ref{sec:two} which similarly holds for this system.

In the following section, we explore a broad class of gauge theories originating from exponential charge symmetries. We show that these theories can constrain quasiparticle mobility and exhibit ground-state degeneracies with UV (lattice size) dependence.

\section{Global and Exponential Charge Symmetries Gauge Theories}\label{sec:two}

Our goal is to establish a class of generalized discrete gauge theories in which charge and flux excitations transform nontrivially under lattice translations. The Fuji–Cheng–Watanabe stabilizer code, perhaps the simplest model in this class, has already demonstrated its potential for revealing constrained dynamics and UV–IR mixing.




For concreteness, we now focus on a special instance of the exponential polynomial symmetries in Eq.~\eqref{exponential} that conserves both the global charge and the exponential charge. As we will argue, this combination is sufficient to give rise to topologically robust quasiparticles with restricted mobility. Concretely, we fix the polynomial function of the lattice points $r=(x,y)$ to be $g_r=1$ and allow $f_r$ to take the values $1$, $x$, $y$, and $x+y$ on a square lattice. This choice yields four independent symmetry generators:
\begin{eqnarray}\label{charges}
G &=& \sum_r  q_r,\nonumber\\
G_x &=& \sum_r a^{x}  \, q_r,\nonumber\\
G_y&=&\sum_r a^{y}  \, q_r ,\nonumber\\ G_{xy} &=& \sum_r a^{x+y}\,  q_r.
\end{eqnarray} 

\begin{figure}[h!]
  \centering
      \includegraphics[scale=0.45]{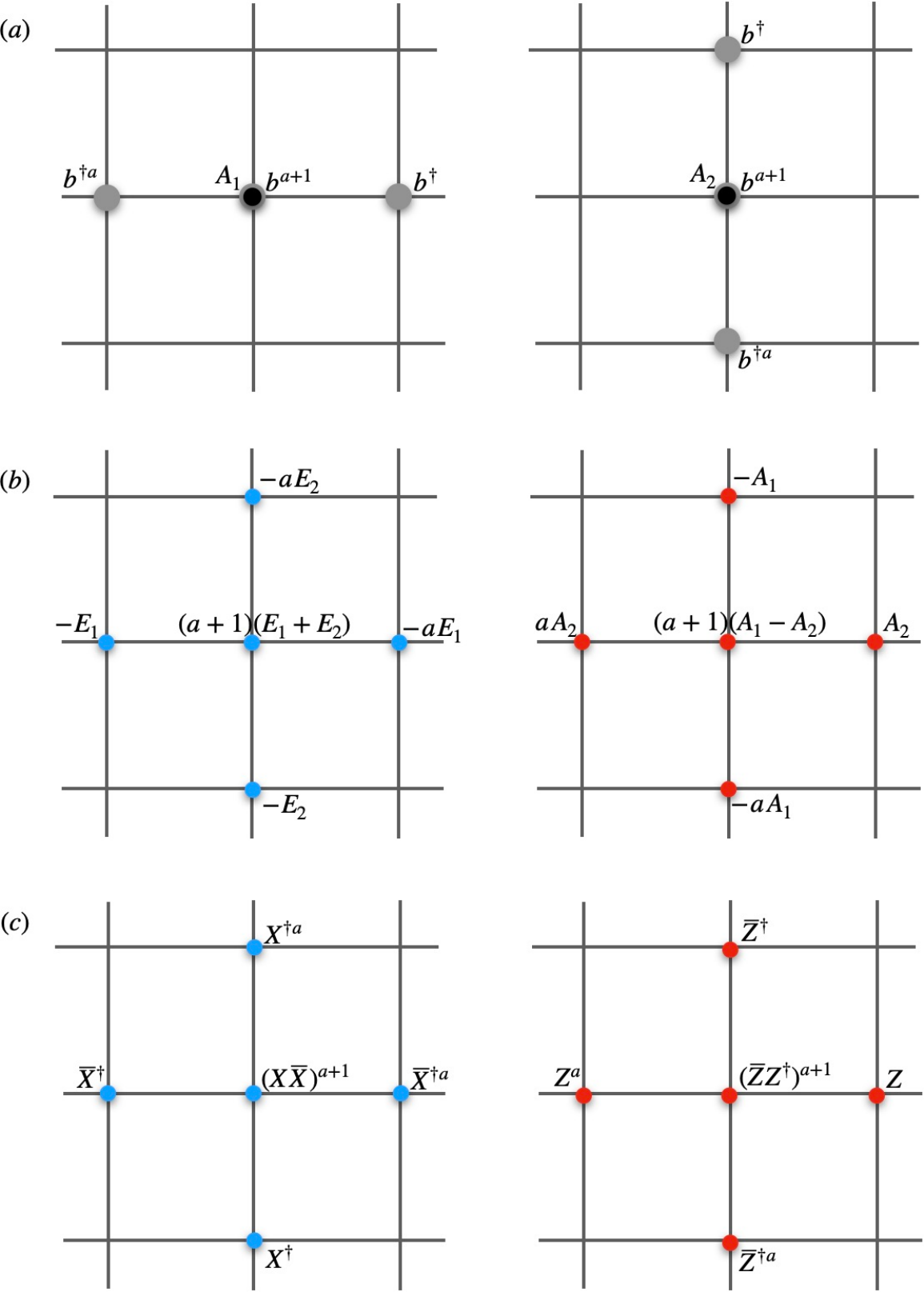}
  \caption{(a) Boson hopping terms, together to the gauge fields, respecting all conserved charges in Eq. \eqref{charges}; (b) Gauss Law and magnetic flux operator in the $U(1)$ gauge theory; (c) Charge and flux operators $\mathcal{Q}_r $ and $\mathcal{B}_r $, respectively, after Higgsing $U(1)$ down to $\mathbb{Z}_N$.} 
\label{exp_symm_model}
\end{figure}

The symmetry defined in Eq.~\eqref{charges} respects both global charge and exponential charge symmetry.
Based on this special conservation law, the possible charge dynamics on the lattice can be written as
\begin{align} 
b^{\dagger a}_r ~b^{a+1}_{r+e_x} ~b_{r+2e_x}^{\dagger}+b_r^{\dagger a } ~b_{r+e_y}^{a+1} ~b_{r+2e_y}^{\dagger}+\text{h.c.}.
\label{co2}
\end{align}
Although other terms in the bosonic Hamiltonian are allowed, such as longer-range hopping terms, they can be generated as higher-order combinations of the boson hopping term in Eq.~\eqref{co2}. We therefore neglect them here, as they do not alter the gauge structure. The cost of imposing the symmetries generated by Eq.~\eqref{charges} is that the boson hopping must occur through collective coordinate moves involving three sites, rather than the usual two-site kinetic terms.

The boson dynamics described in Eq. \eqref{co2} can be interpreted as the creation of $a$ number of $x$-dipoles units (or $y$-dipoles) and the annihilation of a unit charge on the nearby $r+e_x$ (or $r+e_y$) site. Here, a dipole unit refers to a pair of charge and hole separated along the $x$ (or $y$) link. One can explicitly check that these hopping terms respect all the symmetries defined in Eq. \eqref{charges}, and consequently, terms like those in Eq. \eqref{co} are not allowed as they violate the total charge number $G$ conservation. 
In this sense, the exponential charge dynamics are generated at the dipole level.

To elucidate the gauge structure, we gauge the symmetry in Eq.~\eqref{charges} by
coupling the boson field with a gauge potential $A_1,A_2$, as illustrated in Fig.~\ref{exp_symm_model}(a),
\begin{align} 
&(b^{\dagger}_r)^{a} e^{iA_{1,r+e_x}} b_{r+e_x}^{a+1}  b_{r+2e_x}^{\dagger}+(b_r^{\dagger})^{a} e^{iA_{2,r+e_y}} b_{r+e_y}^{a+1}  b_{r+2e_y}^{\dagger}.
\end{align}
The gauge fields $A_1$ and $A_2$ mediate the interaction among the bosons on the three sites, and consequently, live on the vertices of the square lattice, as well as their conjugate partners electric fields $E_1$ and $E_2$. This is to be contrasted to the example in the previous section, where the gauge fields lived on the lattice edges. The gauge potentials are subject to the following transformations
\begin{align} 
&A_{1,r} \rightarrow A_{1,r}+ f_{r+e_x}+a f_{r-e_x}-(a+1)f_r, \nonumber\\
&A_{2,r} \rightarrow A_{2,r}+ f_{r+e_y}+a f_{r-e_y}-(a+1)f_r.
\label{GI}
\end{align}
The generator of these gauge transformations is the generalized Gauss law:
\begin{eqnarray}
    D_1 E_1+D_2 E_2= q_r
\end{eqnarray}
where we defined the ``lattice derivative operators'' as
\begin{eqnarray}
    &D_1f_r=-f_{r+e_x}-a f_{r-e_x}+(a+1)f_r,\nonumber\\
&D_2f_r=-f_{r+e_y}-a f_{r-e_y}+(a+1)f_r.
\end{eqnarray}
It is a straightforward exercise to show that with this Gauss law, the quantities in Eq. \eqref{charges} are automatically conserved in an infinite lattice.

\begin{widetext}
We define the gauge invariant magnetic flux operator, which also lives at the vertices of the square lattice, as illustrated in Fig.~\ref{exp_symm_model}(b).
\begin{align} 
B_r\equiv a A_{2,r-e_x}+A_{2,r+e_x}-(a+1)A_{2,r} -a A_{1,r-e_y}-A_{1,r+e_y}+(a+1)A_{1,r}
\label{flux2}
\end{align}
\end{widetext}

After some algebra, one finds that the magnetic flux operator enforces the following conservation laws on an $L_x \times L_y$ periodic lattice:
\begin{align}
&\sum_r  B_r = 0,\quad
\sum_r a^{L_x-x} \, B_r = 0 ,\nonumber\\
&\sum_r a^{L_y-y}  \, B_r  = 0,\quad
\sum_r a^{L_x+L_y-x-y} \, B_r  = 0.
\label{con2}
\end{align} 
The similarity with the charge conservations in Eq. \eqref{charges} is no coincidence, as the charge and flux operators are dual to each other. As a result, both the total usual flux and the `exponential flux' are conserved.



We now discretize the $U(1)$ gauge theory to $\mathbb{Z}_N$, thereby implementing $\mathbb{Z}_N$ charges. Since the gauge fields reside on the lattice vertices, each site hosts two $\mathbb{Z}_N$ degrees of freedom. We parameterize the field components as $\omega^{E_2}=X,, e^{iA_2}=Z,, \omega^{E_1}=\bar{X},$ and $e^{iA_1}=\bar{Z}$. The resulting gauge theory can then be expressed in terms of a CSS-type code:
\begin{eqnarray}\label{model1}
    H=-\sum_r\mathcal{Q}_r - \sum_r \mathcal{B}_r+\text{h.c.} 
\end{eqnarray}
with $\mathcal{Q}_r$ and $\mathcal{B}_r$ the $\mathbb{Z}_N$ charge and flux operators, defined as

\begin{eqnarray}\label{eq:DCYmodel}
\mathcal{Q}_r  &=& X_r^{a+1} \, \bar{X}^{a+1}_r \, \bar{X}_{r+e_x}^{\dagger a} \, \bar{X}_{r-e_x}^\dagger \, X_{r+e_y}^{\dagger a} \, X_{r-e_y}^\dagger \nonumber\\ \quad \mathcal{B}_r  &=& Z^{\dagger {a+1}}_r \, \bar{Z}_r^{a+1}\, \bar{Z}^\dagger_{r+e_y}\,\bar{Z}_{r-e_y}^{\dagger a}\, Z_{r+e_x}\,  Z_{r-e_x}^{a}
\end{eqnarray}
as illustrated in Fig. \ref{exp_symm_model}(c).

Finally, we emphasize that the aforementioned protocol only works when $a \neq 0 $ mod rad ($N$), so the symmetries we begin with are global symmetries. If $a = p \,\text{rad}(N)$, for some $p\in \mathbb Z$ one can infer that there exists a finite integer $m$ such that $a^{m}=0$ mod $N$. Consequently, the exponential symmetry acts only on a finite number of degrees of freedom for sites $|r|<m$  and some of the symmetries in Eq.~\ref{charges} are reduce to a local symmetry.


\subsubsection{Ground state degeneracy}
We now count the ground-state degeneracy when the system is placed on a torus. Since every term in the Hamiltonian commutes with all others, they can be diagonalized simultaneously. Each operator satisfies $\mathcal Q_r^N=\mathds1$ and $\mathcal B_r^N=\mathds1$, so their eigenvalues are $e^{2\pi i p/N}$ with $p=0, 1, \ldots, N-1$. We focus on the ground-state subspace, corresponding to eigenvalue $+1$ for all operators, which minimizes the Hamiltonian energy:
\begin{equation}
\mathcal{H}_0 = {\ket{\psi}\in \mathcal{H}, \big | , \mathcal{Q}_r \ket{\psi} = +1\ket{\psi} \text{ and }\mathcal{B}_r \ket{\psi} = +1\ket{\psi}}.
\end{equation}
The ground-state space is therefore defined by the local requirement that both the $\mathbb{Z}_N$ charge and the flux vanish everywhere.
In the following, we impose periodic boundary conditions in both $x$ and $y$ directions. These introduce additional global constraints on the allowed field configurations so that the conserved quantities in Eq.\eqref{charges} and Eq.\eqref{con2} remain well defined. In terms of the $\mathbb{Z}_N$ degrees of freedom, the conservation equations in Eq.\eqref{charges} take the form
\begin{eqnarray}\label{conserv}
    \prod_r \mathcal{Q}_r  &=& \mathds{1} \nonumber\\
    \prod_{r } \mathcal{Q}_r ^{\rho_1\, a^{x}}&=& \mathds{1} \nonumber\\
      \prod_{r} \mathcal{Q}_r ^{\rho_2 \,a^{y}}&=& \mathds{1} \nonumber\\
    \prod_{r} \mathcal{Q}_r ^{\rho_{12}\, a^{x+y}}&=& \mathds{1}\label{constraints2}
\end{eqnarray}
and similarly for the flux conservation in Eq.\eqref{con2}, expressed in terms of the $\mathcal{B}_r$ operators. Here, $N_a$ is the greatest divisor of $N$ that is coprime to $a$, and the $\rho$ factors are given by
\begin{eqnarray}
\rho_1 &=& \dfrac{N_a}{\gcd(N_a,a^{L_x}-1)}\nonumber\\
    \rho_2 &=& \dfrac{N_a}{\gcd(N_a,a^{L_y}-1)}\nonumber\\
    \rho_{12} &=& \dfrac{N_a}{\gcd(N_a,a^{L_x}-1,a^{L_y}-1)}
\end{eqnarray}
where $L_x$ and $L_y$ denote the linear dimensions of the lattice in the $x$ and $y$ directions. A detailed derivation of Eq.~\eqref{conserv} and the $\rho$ factors is given in Appendix~\ref{global_constraints}.

The ground-state degeneracy is then
\begin{eqnarray}
    \dim \mathcal{H}_0 = \left[N \gcd(N_a,a^{L_x}-1)\gcd(N_a,a^{L_y}-1)\right.\nonumber\\
    \left.\gcd(N_a,a^{L_x}-1,a^{L_y}-1)\right]^2\label{gsd}
\end{eqnarray}
which depends sensitively on the system size in an oscillatory fashion, as discussed in Appendix~\ref{global_constraints}. This functional dependence is markedly different from that of conventional 3D fracton codes~\cite{Chamon2005-fc,Haah2011-ny,vijay2016fracton,shirley2018fractional}, which exhibit a subextensive ground-state degeneracy. In contrast, our degeneracy oscillates with system size but is bounded above by $N^8$.

\subsubsection{Wilson Loops and Holonomies}\label{holonomies}

In gauge theories, holonomies associated with Wilson line operators reveal the global flux sectors of the ground-state manifold on a torus and define the underlying Wilson algebra. In this section, we derive the Wilson operators \cite{oh2023aspects} for the model in Eq.~\eqref{model1}, based on its gauge structure.

%
%

\begin{figure}[h!]
   \centering
   \includegraphics[scale=0.45]{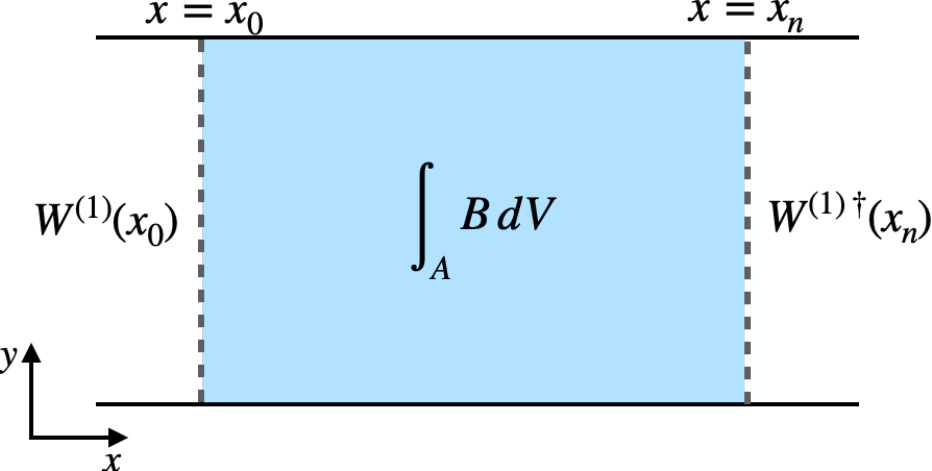}
    \caption{Total flux on an open area $\mathcal A$ (depicted in blue) reduces to string operators acting at the boundaries $x=x_0$ and $x=x_n$. 
    }
   \label{flux_area}
\end{figure}

To obtain the Wilson algebra, we follow the protocol developed in Ref.~\cite{oh2023aspects} as applied to our model. Deriving the Wilson operators requires identifying the generalized Gauss law and Stokes theorem that relate operators defined on a closed surface to the total charge or flux enclosed within that surface.
To this end, we revisit the magnetic and electric conservation laws in Eqs.~\eqref{charges} and \eqref{con2}, evaluated over a closed area as illustrated in Fig.~\ref{flux_area}. The magnetic conservation laws lead to $y$-oriented Wilson line operators that cross the system at a fixed $x$-coordinate. These are given by (for a detailed derivation, see Appendix~\ref{app_wilson}):
\begin{eqnarray}
W^{(1)}(x) &=&  \prod_{i}  Z^a{(x,i)}\, Z^{\dagger}{(x+1,i)} , \nonumber\\
W^{(2)}(x) &=&  \prod_{i}  Z(x,i)\, Z^{\dagger}(x+1,i),\nonumber\\
W^{(3)}(x)  &=&  \prod_{i}  Z^{a^{L_y-i+1}}(x,i)\, Z^{\dagger a^{L_y-i}}(x+1,i) ,  \nonumber\\
W^{(4)}(x)  &=&  \prod_{i}  Z^{a^{L_y-i}}(x,i)\, Z^{\dagger a^{L_y-i}}(x+1,i) .
\end{eqnarray}

Here we clarify our new notation for Pauli operators. Previously we express Pauli operators as $X_l, Z_l$ with lower index $l=(x,y)$ to denote the operator at position $l$. Throughout the discussion of Wilson operators and excitations, we will instead write $X(x,y)$ and $Z(x,y)$ for the Pauli operators at position $l=(x,y)$.
$$
X_l\rightarrow X(x,y), ~~Z_l\rightarrow Z(x,y),
$$
This explicit two-coordinate notation reduces ambiguity because the Wilson line operators are highly anisotropic and direction-dependent.

The $x$-oriented Wilson loops $W^{(i)}(y)$ for $i = 1, \ldots, 4$, which cross the system at coordinate $y$, are defined by similar expressions (see Appendix~\ref{app_wilson}). Throughout this discussion, we use the coordinate dependence to indicate the orientation of the string operators: $W(x)$ denotes $y$-oriented strings, while $W(y)$ denotes $x$-oriented ones.
The dual Wilson lines, corresponding to the electric charge conservation laws in Eq.~\eqref{charges}, are
\begin{eqnarray}
V^{(1)}(y)  &=&  \prod_{i}  X^\dagger(i,y)\, X^{a} (i,y+1), \nonumber\\ 
V^{(2)}(y)    &=&  \prod_{i}  X^\dagger(i,y) \, X(i,y+1), \nonumber\\
V^{(3)}(y)    &=&  \prod_{i}  X^{\dagger a^{i}}(i,y)\, X^{a^{i+1}}(i,y+1) , \nonumber\\ 
V^{(4)}(y)    &=&  \prod_{i}  X^{\dagger a^{i}}(i,y)\,  X^{a^{i}}(i,y+1).
\end{eqnarray}
And same rule applies for the $y$-oriented dual loops $V^{(i)}(x) $. It is worth noting that all the $W^{(i)}$ and $V^{(i)}$ operators have support on closed, double parallel strings. These operators are symmetries of the Hamiltonian and form two copies of a Wilson line algebra, spanning a subspace of degenerate states in the Hilbert space that are locally indistinguishable. Importantly, not all $W^{(i)}$ and $V^{(i)}$ operators commute with each other, due to their non-vanishing intersections. The resulting nontrivial algebra between them, along with their action on the Hilbert space, guarantees ground-state degeneracy-and, more generally, topological degeneracy across all energy sectors.

For later purposes, it is useful to note that all the $W^{(i)}(x)$ and $V^{(i)}(y)$ operators can be constructed from the single-line uniform and exponentially weighted strings:
\begin{eqnarray}
    F^{(1)}(x) = \prod_i Z(x,i),&\quad& F^{(2)}(x) = \prod_i Z(x,i)^{a^{L_y-i}},\nonumber\\
    G^{(1)}(y) = \prod_i X(i,y),&\quad&  G^{(2)}(y) = \prod_i X(i,y)^{a^i}.
    \label{wil}
\end{eqnarray}
In order to specify the Wilson line algebra and recover the ground state degeneracy in Eq. \eqref{gsd}, it is useful to choose specific combinations of the line operators in Eq. \eqref{wil}, which carry the same amount of information as $\{W^{(i)},V^{(i)}\}$.
\begin{eqnarray} 
	S^{(1)}(y)&\equiv& G^{(1)}(y), \nonumber\\
	S^{(2)}(y)&\equiv& G^{(1)}(y)\, G^{( 1)\dagger}(y+1),\nonumber\\
		S^{(3)}(y)&\equiv& G^{(2)}(y)\, G^{( 1)\dagger}(y),\nonumber\\
			S^{(4)}(y)&\equiv& S^{(3)}(y)\,S^{  (3)\dagger}(y+1),\nonumber\\
	\end{eqnarray}
 and 
 \begin{eqnarray}
	T^{(1)}(x)&\equiv& F^{(1)}(x),\nonumber\\
	T^{(2)}(x)&\equiv& F^{(2)}(x)\,F^{ (1)\dagger a^{L_y}}(x),\nonumber\\
	T^{(3)}(x)&\equiv& F^{(1)}(x)\,F^{ (1)\dagger}(x+1),\nonumber\\
	T^{(4)}(x)&\equiv& T^{(2)}(x)\, T^{ (2)\dagger}(x+1).
\end{eqnarray}
The set of operators $\{S^{(i)}, T^{(i)}\}$ is useful as it spans and enumerates the different topological sectors of the ground state manifold. First, one can explicitly check that each one of the $S^{(k)}$ operators commute with all others but with their corresponding pair $T^{(k)}$, namely, $[S^{(i)}, T^{(j)}]=0$ unless $i=j$.  Now, we detail the algebra of each pair ($S^{(k)}, T^{(k)}$) for $k=1, \ldots, 4$ and recover $\dim \mathcal{H}_0$ in Eq. \eqref{gsd}.

For $k=1$, $S^{(1)}(y)$ and $ T^{(1)}(x)$ are $\mathbb{Z}_N$ operators regardless of the system size. It is not difficult to demonstrate that these operators follow the Wilson line algebra
\begin{align} 
&S^{(1)}(y) T^{(1)}(x)=\omega T^{(1)}(x) S^{(1)}(y)
\end{align}
and, thus, $(S^{(1)}(y), T^{(1)}(x))$ span an $N-$fold degenerate Hilbert space.

For the rest of operator pairs, one needs extra caution when evaluating their eigenvalues. For $k=2$, $T^{(2)}(x)$ reduces to a $\mathbb{Z}_{\gcd(a^{L_y}-1,N)}$ operator under closed boundary conditions, so its eigenvalues can only change mod ${\gcd(a^{L_y}-1,N)}$. Likewise, while $S^{(2)}(y)$ is well-defined regardless of system size, it obeys the extra constraint $S^{(2)}(y)\, (S^{(2)\dagger}(y+L_y))^{a^{L_y}}=1$. This again, confirms that the $S^{(2)}(y)$ eigenvalues can only change mod ${\gcd(a^{L_y}-1,N)}$. Thus, $(S^{(2)}(y), T^{(2)}(x))$ span a Hilbert space that is $\gcd(a^{L_y}-1,N)$-fold degenerate. Following the same argument for $k=3$, $(S^{(3)}(y) ,T^{(3)}(x))$ also span a $\gcd(a^{L_x}-1,N)$-fold degenerate Hilbert space.

Finally, for $k=4$,  $S^{(4)}(y)$ reduces to a $Z_{\gcd(a^{L_y}-1, N)}$ operator under closed boundary conditions and its eigenvalues can only change mod ${\gcd(a^{L_y}-1,N)}$. Additionally, the constraint $(T^{(4)\dagger}(x))^{a^{L_x}}T^{(4)}(x+L_x)=1$ reduces the eigenvalues of $T^{(4)}(x)$ to elements of $\mathbb{Z}_{\gcd(a^{L_y}-1,a^{L_x}-1,N)}$. Thus, $(S^{(4)}(y), T^{(4)}(x))$ span a $\gcd(a^{L_y}-1,a^{L_x}-1,N)-$fold degenerate Hilbert space. Likewise, Wilson operators and the associated algebras for $\bar X$ and $\bar Z$ are constructed in a similar manner; details appear in Appendix A1.
Putting all these degeneracy factors together, we are able to reconstruct the GSD expression in Eq. \eqref{gsd}.

\subsubsection{Topological Excitations} \label{sec:GSD}
\begin{figure}[h!]
    \centering
    \includegraphics[scale=0.53]{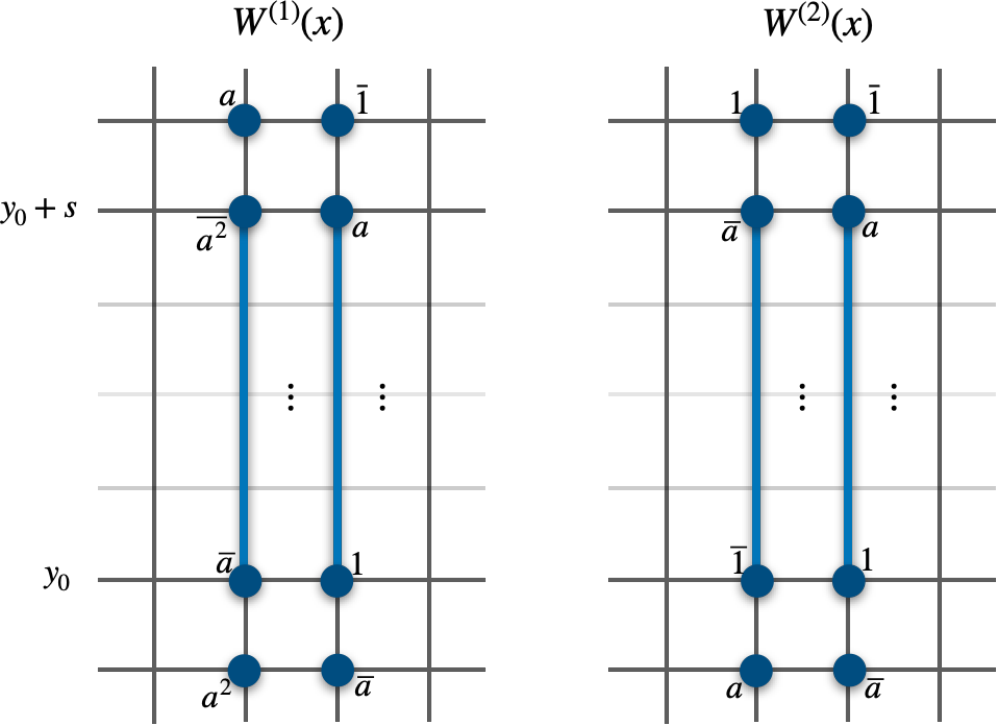}
    \caption{Open $W^{(i)}(x)$ operators extended along y-direction, for $i=1$ and 2, create four quasiparticles bound states of excitations at their endpoints. 
    }
    \label{string_operators}
\end{figure}

We now turn to the study of topological excitations of the Hamiltonian $H$ in Eq.~\eqref{model1}. Owing to the fixed-point nature of $H$, all dispersion bands are flat, and excitations remain localized in real space. Charge and flux quasiparticles are associated with eigenvalues of $\mathcal Q_r$ and $\mathcal B_r$ that differ from $+1$.
As in conventional gauge theories, these excitations appear at the endpoints of open string operators. In our case, we consider open versions of the double-string holonomies $W^{(i)}$ and $V^{(i)}$. A direct inspection shows that all such operators create four-particle bound states at each endpoint. These bound states are fully mobile on the lattice, ensuring that the constituent quasiparticles move collectively.
For illustration, in Fig.~\ref{string_operators}, we display the excitation pattern produced by the open string operators $W^{(1)}(x)$ and $W^{(2)}(x)$ of length $s$ extended along y-direction. As shown, the action of $W^{(i)}$ (and similarly $V^{(i)}$) creates quadrupole bound states localized at the endpoints of the string. 

\subsubsection{Excitations with Constrained Mobility}\label{restricted}

\begin{figure}
    \centering
    \includegraphics[scale=0.53]{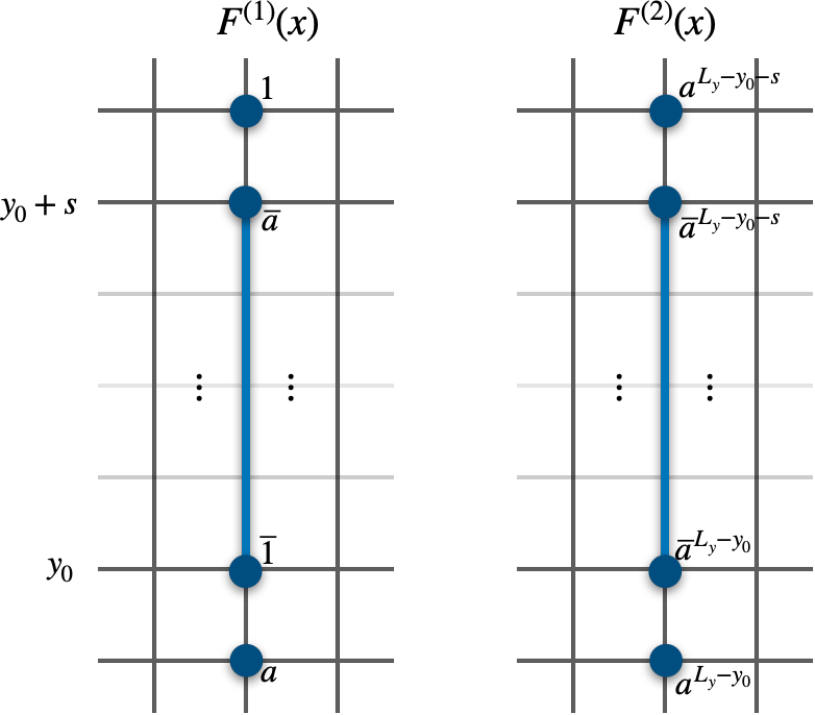}
    \caption{Open $F^{(1)}(x)$ and $F^{(2)}(x)$ operators create two quasiparticles bound states of excitations at their endpoints. The quantum numbers of excitations depend explicitly on the lattice position. 
    }
    \label{dipolar_bound}
\end{figure}

While the open string operators $W^{(i)}$ and $V^{(i)}$ create fully mobile excitations, we now argue that the open strings $F^{(i)}$ and $G^{(i)}$ generate dipolar bound states at their endpoints that behave as lineons—excitations restricted to motion along one-dimensional sublattices. The excitation patterns produced by these operators are illustrated in Fig.~\ref{dipolar_bound}.
The restricted mobility of these dipolar excitations arises from the fact that their associated strings cannot be bent without creating additional quasiparticles from the vacuum. As a result, the only allowed motion of these bound states involves stretching or compressing the $F^{(i)}$ and $G^{(i)}$ strings along straight lines.

\begin{figure}[h!]
    \centering
    \includegraphics[scale=0.53]{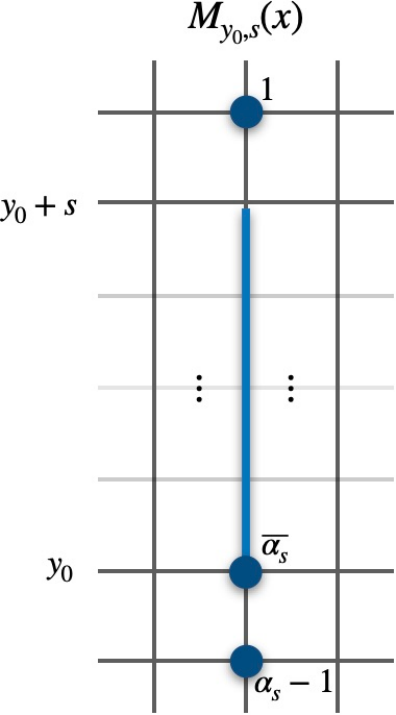}
    \caption{Open $M_{y_0,s}(x)$ operators create one isolated excitation at one of its end-points and two others at the opposite one.}
    \label{Isolated_fract}
\end{figure}

We can also consider operators that move isolated excitations—that is, open string operators that violate only a single $\mathcal{Q}_r$ (or $\mathcal{B}_r$) term at each endpoint. Such operators introduce a characteristic length scale into the system: the minimal step by which a unit charge can be transported (without creating additional excitations that cost finite energy). To analyze this, we define the following sum of integer powers of $a$:
\begin{eqnarray}\label{alpha}
    \alpha_i = \sum_{j=0}^i a^j = \frac{a^{i+1}-1}{a-1}\quad \text{mod }N,
\end{eqnarray}
which can assume integer values mod $N$. Strings of Pauli operators weighted by $\alpha_i$ functions
\begin{eqnarray} \label{strings}
 M_{y_0,s}(x)&=&\prod^s_{i=1}  Z^{\alpha_{s-i}}(x,y_0+i),\nonumber\\
N_{x_0, s}(y)&=&\prod^s_{i=1}  X^{\alpha_i}(x_0+i,y),
\end{eqnarray}
are responsible for creating unit charge at one of their endpoints, as shown in Fig. \ref{Isolated_fract}. As illustrated in the figure, at the other endpoint, however, two excitations are created.

This observation implies that the commensurability between $a$ and $N$ directly impacts the mobility of a single charge. More precisely, let $\ell$ be the smallest integer such that
\begin{eqnarray}\label{emergent_lenght}
    \alpha_\ell = 1\quad \text{mod } N.
\end{eqnarray}
decides mobility condition for this excitation. When the Eq. \eqref{emergent_lenght} has a solution, we can fix $s=\ell$ and the excitation at $(x,y_0-1)$, as shown in Fig. \ref{Isolated_fract}, condenses back to the vacuum. In this case, effectively, $M_{y_0,s}(x)$ has the role of creating a $-1$ and $+1$ charges at $(x,y_0)$ and $(x,y_0+\ell+1)$, respectively. Thus, isolated particles of unit charge can hop with step sizes $\ell$ in both $x$ and $y$ directions and the resulting theory. It is a straightforward exercise to check that for $a=1$, $\alpha_i =i+1$ and the smallest solution to $\alpha_\ell = 1$ is $\ell=N$. For $a=1$ we recover a variation of the $\mathbb{Z}_N$ code studied in Ref. ~\cite{delfino22}, where the emergence of such $N$-sized string operators plays an important role. 

We are interested in the class of theories for which Eq.\eqref{emergent_lenght} does not admit solutions, which can occur when $a$ and $N$ share common prime factors. In such cases, no string operators exist that can move a unit charge excitation between string ends without creating additional excitations. This restricted mobility is reminiscent of higher-rank gauge theories in 3D fracton stabilizer codes\cite{haah2011local,vijay2016fracton}, but it arises from a different origin.
In 3D higher-rank gauge theories, fracton behavior results from subsystem charge conservation, which forbids the motion of an isolated charge in any direction and leads to extensive ground-state degeneracy. In contrast, the restricted mobility in our case can be viewed as translation symmetry–enriched topological order\cite{williamson2019fractonic,pace2022position,delfino2024anyon}. Specifically, due to the conservation law in Eq.\eqref{charges}, charge excitations transform nontrivially under lattice translations:
\begin{eqnarray}\label{chargessymm}
T_x G T^{-1}_x =G \nonumber\\
T_x G_x T^{-1}_x = a G_x .
\end{eqnarray}
Thus, we can view the 2D lattice model as a topological phase whose gauge charges $G_x$ transform nontrivially under lattice translations. When a single $G$-excitation (as defined in Eq.\eqref{charges}) is created, it also carries a gauge charge $G_x$ that depends on position $x$ as $a^x$. 
Such position dependence of the gauge charge also appears in other 2D modulated gauge theories with dipole or quadrupole conservation laws~\cite{delfino22,oh2022rank,delfino2024anyon,pace2022position,oh2023aspects}. The difference is that, in 2D modulated gauge theories, the gauge charge is position dependent and \textit{oscillates periodically} on the lattice, allowing the system to be viewed as a superlattice with distinct topological sectors on different sublattices\cite{pace2022position}. By contrast, in our exponential gauge theory the gauge charge exhibits periodic oscillations only after a preperiodic stage; see Sec.~\ref{sec:mobi} for further discussions.

It is interesting to note that Eq. \eqref{emergent_lenght}, for general $N$ and $a\neq 0$ mod rad($N$),  is a variation of the order-finding problem, a question which is known for being a hard problem to solve in classical computers for large enough $a$ and $N$. To see the connection, we use the closed form for $\alpha_i$ in Eq. \eqref{alpha}, and the mobility condition Eq. \eqref{emergent_lenght} can be rewritten as
\begin{eqnarray}\label{order_finding1}
    a^{\ell+1}-1=(a-1)(1+pN)\quad p\in \mathbb{Z},\nonumber \\
    \Rightarrow a^{\ell+1}=a \quad \text{mod }N.
\end{eqnarray}
Dividing  both sides of the equation by $a$, we get that the mobility condition reads
\begin{eqnarray}\label{order_finding}
    a^\ell = 1 \quad \text{mod }M_a,
\end{eqnarray}
where $M_a \equiv N/\gcd(N,a)$. When $a$ and $N$ are coprime ($M_a=N$), from Euler's totient theorem, there always exists a finite integer $\varphi(N)$ such that $a^{\varphi(N)}-1=0 \pmod{N}$. 
In this case,  Eq. \eqref{order_finding} always admits a solution, $\ell = \varphi(N)$, and the connection to the order-finding problem is explicit (i.e. finding the smallest $r$ such that $a^r=1$ mod $N$). 
    This concept raises an intriguing notion: even when dealing with exactly solvable Hamiltonians, there still exist some properties of the system, in this case, the emergent step size that particles can hop, that can be computationally expensive to determine for large enough $a$ and $N$. 
    

     Although unit charges and fluxes can be completely immobile when Eq. \eqref{emergent_lenght} admits no solutions, single particles with higher charges (or flux) might be able to move. This follows from considering multiple powers, say $\beta$, of $M_{r_0, s}$ or $N_{r_0,s}$ operators on the vacuum. Following similar arguments as before, the condition for charges (or fluxes) of quantum number $\beta$ to move is the existence of an integer $\ell$ such that
   \begin{eqnarray}
       \beta \, \alpha_\ell =1 \quad \text{mod }N,
   \end{eqnarray}
   which is way less restrictive than the condition for unit charges and fluxes in Eq. \eqref{emergent_lenght}. 

\subsection{Discussion on Mobility of the Charge Excitations}\label{sec:mobi}

So far we have examined mobility constraints in various 2D topological orders arising from gauging exponential symmetry. At this stage, it is essential to clarify what we mean by particle immobility and how different definitions in the literature converge and where they differ.

There are two common definitions of particle immobility for translation invariant Hamiltonian:

\textbf{A. String-operator mobility}: A local excitation is mobile if there exist bounded-width string operators that move it to arbitrarily large separations, mapping it to \textit{another local excitation} with finite energy scale. The operator’s support width is bounded by a distance-independent constant.

\textbf{B. Local-translation mobility}: A local excitation is mobile if there exists a local operator that converts the excitation into its own translate, i.e., the action of translation by \textit{several units} can be mimicked by a local operator. In this definition, we only expect the excitation to move by several units, so in the coarse-grained RG sense the particle is mobile. Examples include the Wen plaquette\cite{wen03} model and certain Rank-2 Toric-Code variants\cite{oh2022rank,oh2022effective,pace2022position}, where a charge moves only in multiples of M sites. In these theories, one can treat M-site unit cell as an enlarged superlattice, and the resultant theory is more or less like the usual topological order with different flavor of excitations in the enlarged unit cell.

Based on the definition A, our unit-charge excitation with $\mathcal{Q}=1$ generated by string operators in Eq.~\ref{strings} is mobile. One can move an isolated unit charge $\mathcal{Q}(x,y)=1$ to the end of a string of length $s$. The two sites at the string’s end then carry a bound state with charges $\mathcal{Q}(x,y+s-1)=\alpha_s$ and $\mathcal{Q}(x,y+s)=\alpha_s-1$ as defined in Eq.~\ref{alpha}. In this sense, the string operator not only transports the unit charge but also changes it into another type of excitation: the original unit charge branches into a bound pair separated by one lattice spacing. Nonetheless, this bound excitation should still be regarded as a local excitation.

In the translation-invariant stabilizer model (Eq.~\ref{eq:DCYmodel}), the original unit charge has energy
$$
E_1=\cos\frac{2\pi}{N}.
$$
The bound state with $\mathcal{Q}(x,y+s-1)=\alpha_s$ and $\mathcal{Q}(x,y+s)=\alpha_s-1$ has energy
$$
E_{\text{bound}}=\cos\frac{2\pi(\alpha_s-1)}{N}+\cos\frac{2\pi\alpha_s}{N}.
$$
Despite this energy difference, moving the unit charge incurs only a finite cost, so string operators can convert a unit charge into another type of local excitation with arbitrary large separation. Thus, under definition A, our unit-charge excitation is mobile. We note that definition A is the standard criterion used for many 3D fracton models~\cite{haah2011local,vijay2016fracton}: a unit charge is deemed mobile if a string operator can move it to arbitrarily large separations while the excitation’s support remains local.

Based on definition B, our unit charge excitation is immobile for non-coprime $a$ and $N$. Notably, definition B requires that a unit charge excitation can be converted into its translate (carrying the same energy) via a local operator. 
However, the string operator defined in Eq.~\eqref{strings} not only transports the unit charge but also branches it into a bound pair, which is not simply a translate of the original excitation.

Given that our theory comes from gauging the exponential symmetry defined in Eq.~\eqref{charges}, the unit charge excitation at position $(x, y)$ carries gauge charge $G=1$, $G_x=a^x$, $G_y=a^y$, and $G_{xy}=a^{x+y}$. To realize a local operator that translates the excitation by $l$ sites along the $x$-direction and preserves the same gauge charge, we require $a^l \equiv 1 \pmod N$ so that, after translation, the excitation carries the same set of gauge charges.

However, for non-coprime $a$ and $N$, there is generally no integer $l$ satisfying $a^l \equiv 1 \pmod N$. This also suggests that we cannot coarse-grain the system into a superlattice where the gauge charge oscillates \textit{fully periodically} (without any pre-periodic cycle) within an enlarged unit cell. This arises from the fact that the exponential gauge charge $G_x = a^x$ is a periodic function only after a finite preperiodic cycle.
Consider the residue sequence
$$
G_x \equiv a^x \bmod N,\qquad x=0,1,2,\dots
$$
It is eventually periodic regardless of the choice of a, since there exist integers $\mu\ge 0$ and $\lambda\ge 1$ such that
$$
G_{x+\lambda}\equiv G_x \bmod N \quad \text{for all } x\ge \mu .
$$
The terms with $x<\mu$ form the \textit{preperiodic cycle}; they would not appear in the later cycle (in particular, $a^0=1$ may never recur).

Write $N=\prod p_i^{k_i}$ and split
$$
N_0=\prod_{p_i\mid a} p_i^{k_i},\qquad N_1=\prod_{p_i\nmid a} p_i^{k_i}.
$$
For each $p\mid a$ with $p^{k}\parallel N$ and $v_p(a)$ the $p$-adic valuation, once
$$
\mu=\max_{p^{k}\parallel N,\ p\mid a}\left\lceil \frac{k}{v_p(a)}\right\rceil
$$
we have $a^x\equiv 0 \pmod{p^{k}}$ for all $x\ge \mu$. Hence after $x\ge\mu$ the sequence reduces modulo $N_1$, where $a$ is a unit and
$$
G_{x+\lambda}\equiv G_x \pmod{N_1},\qquad
\lambda \mid \operatorname{lcm}_{p^{k}\parallel N_1}\lambda(p^{k}),
$$
with $\lambda(\cdot)$ the Carmichael function. Consequently, $G_x$ is finally periodic, while the initial values may belong to the preperiod and never reappear.

A simple example:
$$
a=2,\ N=12:\quad 1,2,4,8,4,8,\dots
$$
Here $x=0,1$ are preperiodic; from $x=2$ onward the sequence is periodic. Thus, the charge with unit $\mathcal{Q}=1,a$ are immobile under definition B.


Finally, we note that since the gauge charge $G_x = a^x$ is eventually periodic, certain multiple unit charge excitations are always mobile. In particular, if we consider the charge excitation $\mathcal{Q} = a^{\mu}$ (where $\mu$ is the preperiod length), such an excitation is always mobile under definition B (and also mobile under definition A). Notably, with appropriate boundary conditions or system size, these excitations can be translated along a non-contractible loop, forming holonomy operators as defined in Eq.~\eqref{strings}. These holonomies are responsible for the ground state degeneracy of the system.

In summary, the unit-charge excitation in our exponential gauge theory could be immobile (when $a$ and $N$ are non-coprime) in the sense that no local operator maps it to its own translate (a translation by a finite number of sites). However, string operators can move a unit charge and, in doing so, branch it into a bound pair of charges; these products should still be regarded as local excitations.

Through our discussions, the exponential gauge theory can be viewed as a translation symmetry-enriched topological order in which lattice translations permute (or transmute) distinct gauge charges. 
More broadly, the connection between fracton gauge theories, whose quasiparticles exhibit mobility constraints, and symmetry-enriched topological order has been explored in Ref.~\cite{williamson2019fractonic}. In that framework, one start with a toric code enriched by a global symmetry $G$. Transporting an anyon along a spatial path changes the total $G$-charge carried by the system; equivalently, the string operators that move anyons are charged under $G$. After gauging $G$, the resulting gauge theory can develop fractonic features, in the sense that certain excitations acquire restricted mobility.
Motivated by this perspective, we propose an alternative route to our exponential gauge theory. One can start from the standard 2d $\mathbb{Z}_N$ toric code, and enrich the charge excitation by a spatially modulated symmetry $G$. Upon gauging $G$, the resulting theory becomes closely reminiscent of the exponential gauge theory. We elaborate on this construction in Appendix~\ref{sec:appc}.

\section{2D exponential symmetries and Beyond}\label{beyond}

In this section, we explore other variations of models with exponential symmetries. Section~\ref{cs} examines the charge–flux attachment mechanism applied to previously studied models to construct non-CSS codes. Section~\ref{other} presents lattice models invariant under dipolar exponential symmetries. Finally, Section~\ref{subsystem} discusses a 2D symmetry-protected topological phase protected by exponential symmetries.

\subsection{Non-CSS Codes: Chern-Simons-like Theories}\label{cs}
At this stage, we have developed a taxonomy of generalized discrete gauge theories from \textit{exponential polynomial symmetries}. These generalized electromagnetic theories can be treated as CSS stabilizer codes, with ground states obeying zero charge and flux conditions. In this section, our aim is to create non-CSS codes exhibiting similar exponential charge conservation. Intriguingly, these theories can be manifested as Chern-Simons theories, where charge and flux are bonded together.

To set the stage, we begin with a modified Gauss law that features charge structures similar to the ones in Eq.~\eqref{charges}, but with a modification: unit charges are bounded  to unit gauge fluxes. This flux-charge binding process has the potential to give rise to a Chern-Simons type gauge theory~\cite{you2019fractonic,delfino22} in contrast to a Higgsed Maxwell theory.
In conventional Maxwell-type gauge theories, the ground-state manifold is obtained by projecting the local Hilbert space onto sectors with vanishing net charge and flux. In the presence of flux–charge binding, the net-charge condition is automatically satisfied whenever the theory is flux-free. Moreover, any electric charge excitation necessarily carries gauge flux, and vice versa, resulting in fractional statistics between charges and fluxes.

\begin{figure}[h!]
  \centering
      \includegraphics[scale=0.42]{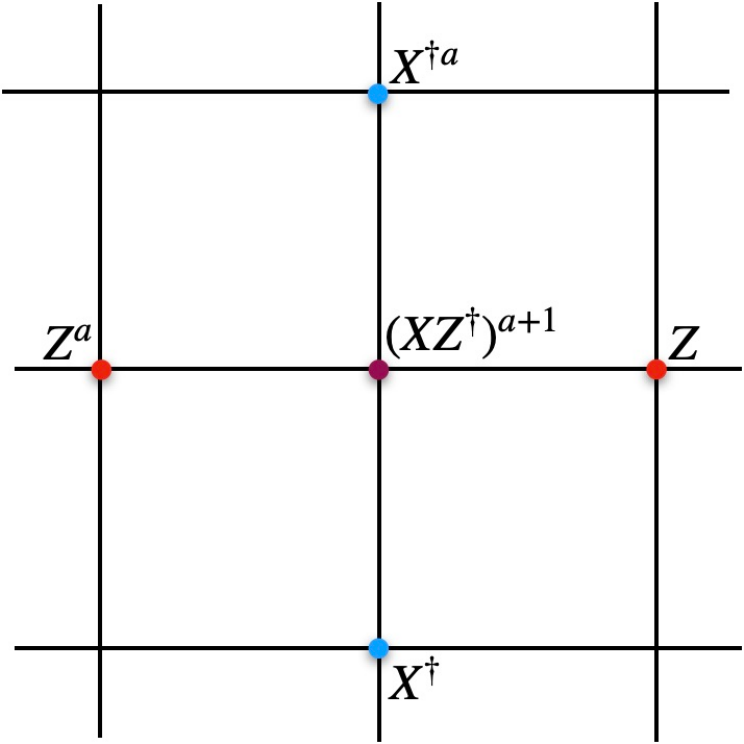}
  \caption{$\mathbb{Z}_N$ exponential symmetry operators under charge and flux attachment.} 
\label{non_css}
\end{figure}

To create a Chern-Simons-type coupling, we assign the charge density to be equal to the local flux,
\begin{align} 
q_r =D_1 E_1+D_2 E_2\stackrel{!}{=}B_r
\end{align}
where $B_r$ is defined in Eq. \eqref{flux2}. 
A sufficient solution to this equation is to
impose a local projection between the two sets of $\mathbb{Z}_N$ Pauli operators $X,Z$ and $\overline X,\overline Z$, such that
\begin{align}\label{localmap}
\bar{Z}=X^a,\quad  \bar{X}=Z^{\dagger a}, \quad Z=\bar{X}^{\dagger a}, \quad X=\bar{Z}^{a}
\end{align}
It is worth noticing that Eq. \eqref{localmap} renders a self-consistent solution only if $a^2=1\pmod{N}$. Under this condition, the local Hilbert space per site is reduced from $N^2$ to $N$, with only one set of Pauli operators per site. The Hamiltonian in Eq.~\eqref{model1} is reduced to,
\begin{align} 
H=-\sum_r\left(X_r^{a+1} \,Z^{\dagger{a+1}}_r\, Z_{r+e_x} \,Z_{r-e_x}^{a}\, X_{r+e_y}^{\dagger a} \,X_{r-e_y}^{\dagger} +\text{h.c.}\right)
\end{align}
which is depicted in Fig. \ref{non_css}.

Now, there is only one stabilizer per site that projects both the charge and the flux to the null sector. As the constraint of flux and charge now becomes essentially the same thing, following the discussion in Sec.~\ref{sec:GSD}, it is not difficult to see that the ground state degeneracy is reduced to 
\begin{eqnarray}
      \text{GSD}_{\text{CS}} = N \gcd(N_a,a^{L_x}-1)\gcd(N_a,a^{L_y}-1) \nonumber\\
    \gcd(N_a,a^{L_x}-1,a^{L_y}-1).
\end{eqnarray}
It is worth noting that the solution given in Eq.~\eqref{localmap} is not unique. Rather, it is a sufficient and straightforward solution that works only if $a^2=1\pmod{N}$. It is possible that there are other solutions that satisfy the flux-charge binding constraint, but we will defer further investigation of them, as well as effective theories of such models, to future research.

\subsection{Other Generalizations}\label{other}

We extend our discussion by introducing an alternative charge arrangement in Eq. \eqref{exponential}, taking $g_r=1,x,y,xy$ and allowing $f_r$ to take on the values of $x+y$. While we do not plan to study the resulting theories in detail, we introduce the basic setup and general ideas, furnishing an example of an even more exotic exponential polynomial symmetry. The symmetry generators are explicitly given by
\begin{eqnarray} 
G &=& \sum_r  a^{x+y} \,q_r,\nonumber\\
G_x &=& \sum_r x\, a^{x+y} \,q_r,\nonumber\\
G_y &=& \sum_r y\, a^{x+y} \,q_r,\nonumber\\ G_{xy}&=& \sum_r xy \,a^{x+y} \,q_r.
\label{con3}
\end{eqnarray}

\begin{figure}[h!]
  \centering
      \includegraphics[scale=0.30]{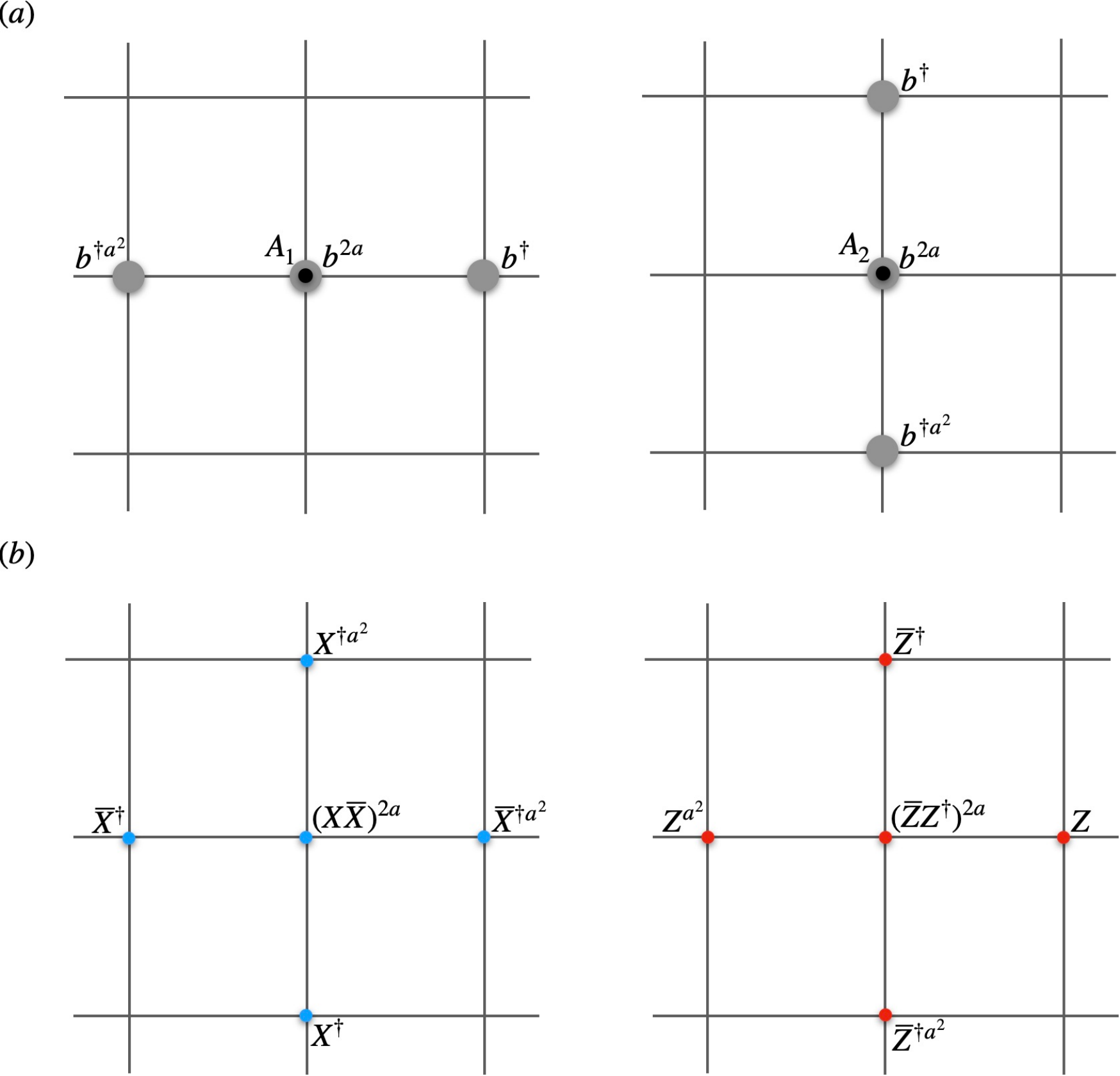}
  \caption{(a) Boson hopping structure, respecting all conserved quantities in Eq. \eqref{con3}; (b) $\mathbb{Z}_N$ charge and flux operators after gauging the exponential symmetries.} 
\label{other_generalization}
\end{figure}

The symmetries defined in Eq.~\eqref{con3} respect exponential charge, dipole, and quadrupole symmetries.
The following Hamiltonian is a possible bosonic lattice realization of such conservation laws,
\begin{align} 
b_r^{\dagger a^2}\, b^{2a}_{r+e_x}\, b_{r+2e_x}^{\dagger}+b_r^{\dagger a^2} \, b^{2a}_{r+e_y} \, b_{r+2e_y}^{\dagger}+\text{h.c.}.
\end{align}
The gauge structure, as before, can be studied by gauging the symmetries generated by Eq.~\eqref{con3}. We
couple the charged particles with a gauge potential $A_1,A_2$ 
\begin{align} 
b_r^{\dagger a^2} b^{2a}_{r+e_x}e^{iA_{1,r+e_x}} b_{r+2e_x}^{\dagger}+b_r^{\dagger a^2} b^{2a}_{r+e_y}e^{iA_{2,r+e_y}} b_{r+2e_y}^{\dagger}
\end{align}
as illustrated as Fig.~\ref{other_generalization}(a).

The  $A_1$ and $A_2$ fields live on the sites of the square lattice as the conjugate partners of the electric field $E_1$ and $E_2$, respectively, and are subject to gauge transformations given by:
\begin{align} 
&A_1 \rightarrow A_1+ f_{r+e_x}+a^2 f_{r-e_x}-2a\,f_r\nonumber\\
&A_2 \rightarrow A_2+ f_{r+e_y}+a^2  f_{r-e_y}-2a\,f_{r}                \end{align}
We hereby define the gauge invariant magnetic flux operator as
\begin{eqnarray}
B_r&=&a^2 A_{2,r-e_x}+A_{2,r+e_x}-2a\,A_{2,r} \nonumber\\
&-&a^2 A_{1,r-e_y}-A_{1,r+e_y}+2a\,A_{1,r}.
\label{flux3}
\end{eqnarray}

The gauge transformations and the gauge flux above define a $U(1)$ gauge theory, which likely to be confined in 2+1D. Since we are interested in deconfined phases we, again, discretize the $U(1)$ gauge group down to $\mathbb{Z}_N$. There are two $\mathbb{Z}_N$ degrees of freedom at each vertex of the square lattice and the resulting gauge theory can be written as
\begin{eqnarray} 
H&=&-\sum_r\left(X^{2a}_r\, \bar{X}_r^{2a} \, \bar{X}_{r+e_x}^{\dagger a^2}\, \bar{X}^\dagger_{r-e_x}\, X_{r+e_y}^{\dagger a^2} \, X_{r-e_y}^{\dagger}\right.\nonumber\\
&\quad&+\left.Z^{\dagger 2a}_r \, \bar{Z}_r^{2a} \, \bar{Z}_{r+e_y}^{\dagger} \, \bar{Z}_{r-e_y}^{\dagger a^2}\,  Z_{r+e_x} \, Z_{r-e_x}^{a^2}+\text{h.c.}\right)
\end{eqnarray}

The low-energy physics of this model can be analyzed in the same manner as described in Sec.~\ref{sec:two}, which we do not repeat here. In general, it is expected to behave similarly to the Hamiltonian in Eq.~\eqref{model1}, differing only in minor aspects such as the explicit form of the ground-state degeneracy and the configurations of particle–bound states created at the endpoints of open string operators.

\subsection{Subsystem Exponential Symmetry in 2D: Spontaneous Symmetry Breaking and Symmetry Protected Topological Order}\label{subsystem}

\begin{figure}[h!]
  \centering
      \includegraphics[scale=0.42]{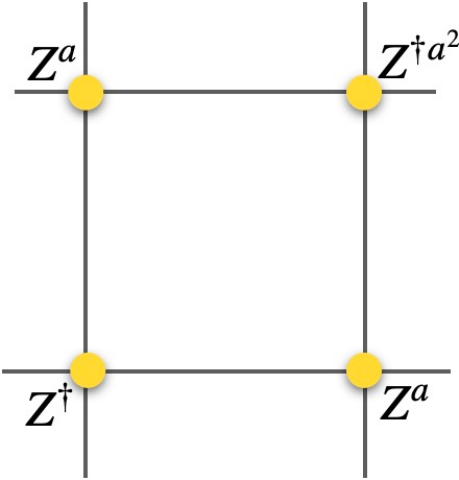}
  \caption{Subsystem exponential symmetry stabilizer $\mathbb{Z}_N$ model, defined on a square lattice.} 
\label{subsystem_exp}
\end{figure}
So far, we have discussed the concept of exponential polynomial symmetries, characterized by charge operators with exponentially modulated spatial variation \cite{Watanabe2022,hu2023spontaneous}. Gauging these symmetries yields gauge theories that can be regarded as spatial symmetry–enriched topological orders, featuring ground-state degeneracy that oscillates with system size and quasiparticles with constrained mobility.
Although we will not go into extensive detail, in this section we introduce a novel type of symmetry, which we term subsystem exponential symmetry. This symmetry is defined by a charge operator with exponential spatial modulation that is conserved on submanifolds encompassing any set of $x$-rows and $y$-columns.
\begin{align}
    U^x(y)=\prod_{m} X^{a^m}_{r+m e_x}, \quad U^y(x)=\prod_{m} X^{a^m}_{r+m e_y}.
    \label{expsub}
\end{align}
The operators for such symmetry are reminiscent of subsystem symmetries, with the difference that the charge operator in each sub-manifold exhibits spatial modulation.

One can build a series of intriguing phases that exhibit various subsystem exponential symmetry patterns.  A typical phase is the spontaneous symmetry-breaking phase leveraged by the following clock Hamiltonian on the 2D square lattice, 
\begin{align}
 &H=-\sum_r\left(Z^{\dagger}_r\, Z^{a}_{r+e_x} Z^{a}_{r+e_y} Z^{\dagger a^2}_{r+e_x+e_y}+\text{h.c.}\right)
\end{align}

The Hamiltonian terms are also shown in Fig.~\ref{subsystem_exp} and  exhibit the symmetry refined in Eq.~\eqref{expsub}. A classical pattern of ground states of this Hamiltonian can be determined by arbitrarily choosing an orientation for the spins along a specific x-row and a specific y-column. This suggests that the ground state entropy scales sub-extensively with the system size $L_x+L_y-1$. The ground state pattern possesses long-range correlation, characterized by a non-vanishing four-point correlation function,
\begin{align}
 &G(p,q)=\langle Z^{\dagger }_r \,Z^{a^q}_{r+qe_x} Z^{a^p}_{r+pe_y} Z^{\dagger a^{p+q}}_{r+qe_x+pe_y} \rangle.
\end{align}
Notably, this model reduces to the plaquette Ising Hamiltonian, explored in Ref.~\cite{Xu2004-oj} for $a = 1$. 

\begin{figure}[h!]
  \centering
      \includegraphics[scale=0.29]{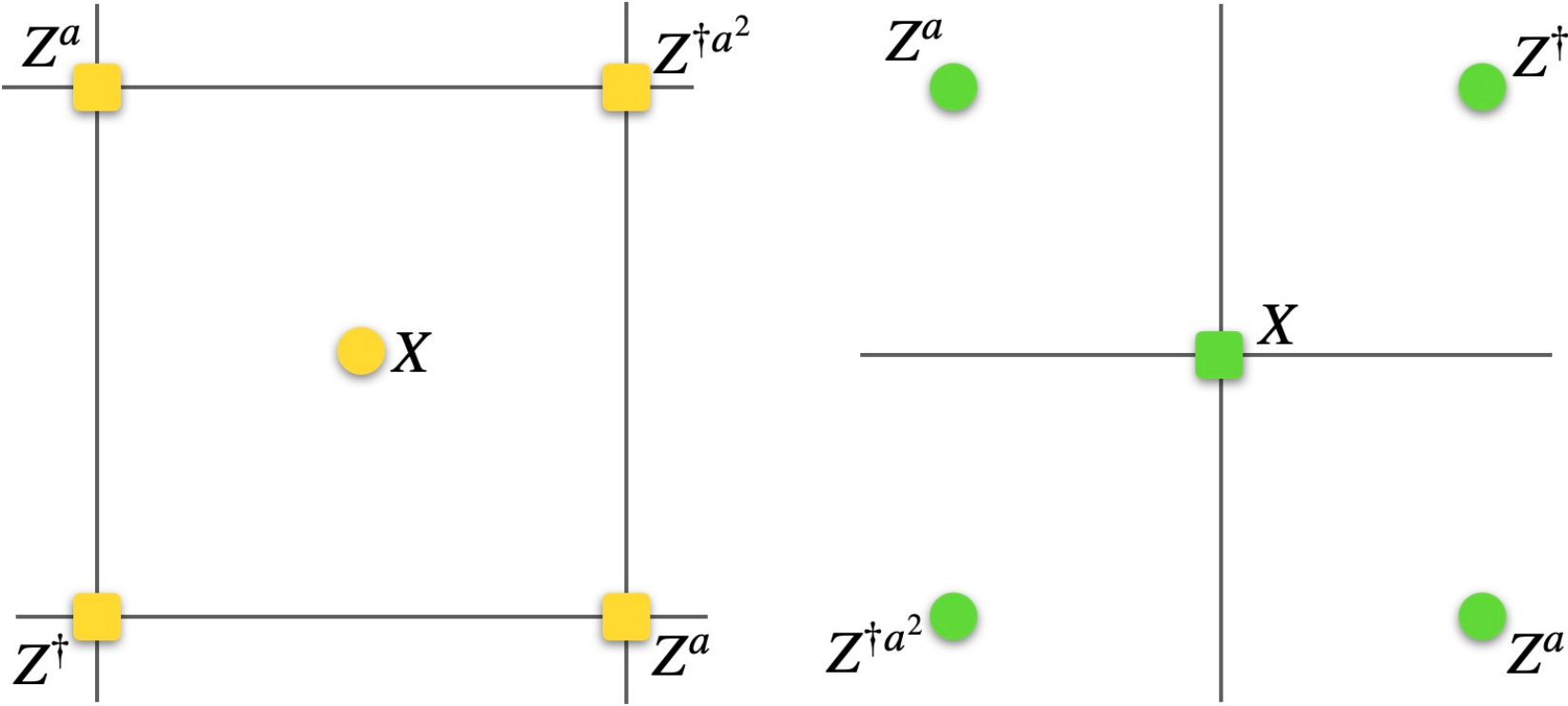}
  \caption{Operators defined on the decorated square lattice centered at plaquettes $p$ and vertices $r$, respectively.} 
\label{decorated_plaquette}
\end{figure}

Symmetry and quantum entanglement come together and give rise to phases of matter nowadays known as Symmetry Protected Topological (SPT) phases. When the exponential subsystem symmetry we defined in Eq.~\eqref{expsub} remains unbroken, the symmetry itself can still enrich quantum correlation, leading to a symmetry-protected topological order state with non-local quantum ordering and protected boundary modes. Here, we introduce a simple, exactly solvable Hamiltonian whose ground state achieves such SPT order. Consider a decorated square lattice, where $\mathbb{Z}_N$ degrees of freedom lie at both vertices $r$ and plaquette centers $p$. We propose the following fixed-point Hamiltonian, also illustrated in Fig. \ref{decorated_plaquette}.
\begin{eqnarray}
 H&=&-\sum_r\left(Z^{\dagger}_r\, Z^{a}_{r+e_x} Z^{a}_{r+e_y} Z^{\dagger a^2}_{r+e_x+e_y} {X}_{r+\frac{e_x}{2}+\frac{e_y}{2}}\right. \nonumber\\
 &\quad&+\left.{Z}^{\dagger a^2}_{p} \,{Z}^{a}_{p+e_x} {Z}^{a}_{p+e_y} {Z}^{\dagger }_{p+e_x+e_y} X_{p+\frac{e_x}{2}+\frac{e_y}{2}} +\text{h.c.}\right)\nonumber
\end{eqnarray}

The spins on the vertex site $r$ exhibit a set of exponential subsystem symmetries (denoted as $P^x$, $P^y$), and the same applies to the spins on the plaquette-centered sites $p$ (denoted as $R^x$, $R^y$),
\begin{align}
    &P^x(y)=\prod_{m} X_{r+m e_x}^{a^m}, \qquad  P^y(x)=\prod_{m} X_{r+me_y}^{a^m},\nonumber\\
    &R^x(y)=\prod_{m} {X}_{p+me_x}^{a^{(L-m)}}   , \qquad R^y(x)=\prod_{m} {X}_{p+me_y}^{a^{(L-m)}}.
    \label{expsub2}
\end{align}
Here $L$ denotes the system size.
Although we will not conduct a comprehensive study of this model, there are a few key points that signify its novelty. The Hamiltonian features a decorated defect structure, characterized by the frustrated plaquette Ising coupling between the four spins on the plaquette corners, decorated with a charge at the center. Consequently, the ground state displays non-vanishing membrane order
\begin{align}
G(p,q)= \langle &Z^{\dagger}_r Z^{a^q}_{r+qe_x} Z^{a^p}_{r+pe_y} Z^{\dagger a^{p+q}}_{r+qe_x+pe_y} \nonumber\\
 &\prod^{q}_{i=1} \prod^{p}_{j=1} {X}^{a^{i+j}}_{r+(i-1/2)e_x+(j-1/2)e_y}\rangle
\end{align}
The quantity $G(p,q)$ can be perceived as a non-local correlation where the four-point correlator among the spins located at the vertices is locked with the total exponential charge of the spins residing at the center of the plaquette inside. 
Intriguingly, this non-vanishing membrane order implies that the SPT wave function, albeit can be prepared by a finite depth local unitary circuit, contains spurious long-range topological entanglement entropy, as proposed in Ref.~\cite{williamson2019spurious}.

\section{3D Fracton Topological Order from Gauging Exponential Symmetry}
In this section, we extend the concept of gauging exponential symmetry to 3D and intertwine it with subsystem symmetry. In particular, we construct a family of stabilizer codes in which anyon excitations exhibit restricted mobility due to the conservation of subsystem exponential symmetry. These codes can be viewed as a generalization of the type-I fracton models introduced in Ref.~\cite{vijay2016fracton}.

\subsection{Exponential symmetry decoration of Planeon Lineon code}

We first consider an anisotropic fracton model that incorporates the exponential charge structure into the Planeon–Lineon model proposed in Ref.~\cite{shirley2018fractional}. Our construction is based on a specific exponential charge conservation law defined over all $xz$ and $yz$ planes:
\begin{eqnarray}\label{3dcharges}
\sum_{r \in \text{yz-plane}} a^{y} \, q_r=G_{zy}(x),\quad 
\sum_{r \in \text{xz-plane}} a^{x} \, q_r=G_{xz}(y).\nonumber\\
\end{eqnarray} 
Here the subscript of $q_r$ indicates a charge located at site $r$. $G_{zy}(x)$ denotes the subsystem symmetry that measures the exponential charge on the $yz$-plane at fixed $x$.
Based on this special conservation law, the possible charge dynamics on the lattice can be written as
\begin{align} 
b_r (b_{r+e_x}^{\dagger})^a (b_{r+e_y}^{\dagger})^a b_{r+e_y+e_x}^{a^2}+b_r b_{r+e_z}^{\dagger}+\text{h.c.}
\end{align}
Which contains a plaquette interaction on the x-y plane and a boson hopping term along the z-direction that respects
the subsystem special exponential charge conservation defined in Eq.~\eqref{3dcharges}. 

\begin{figure}[h]
    \centering
   \includegraphics[width=0.5\textwidth]{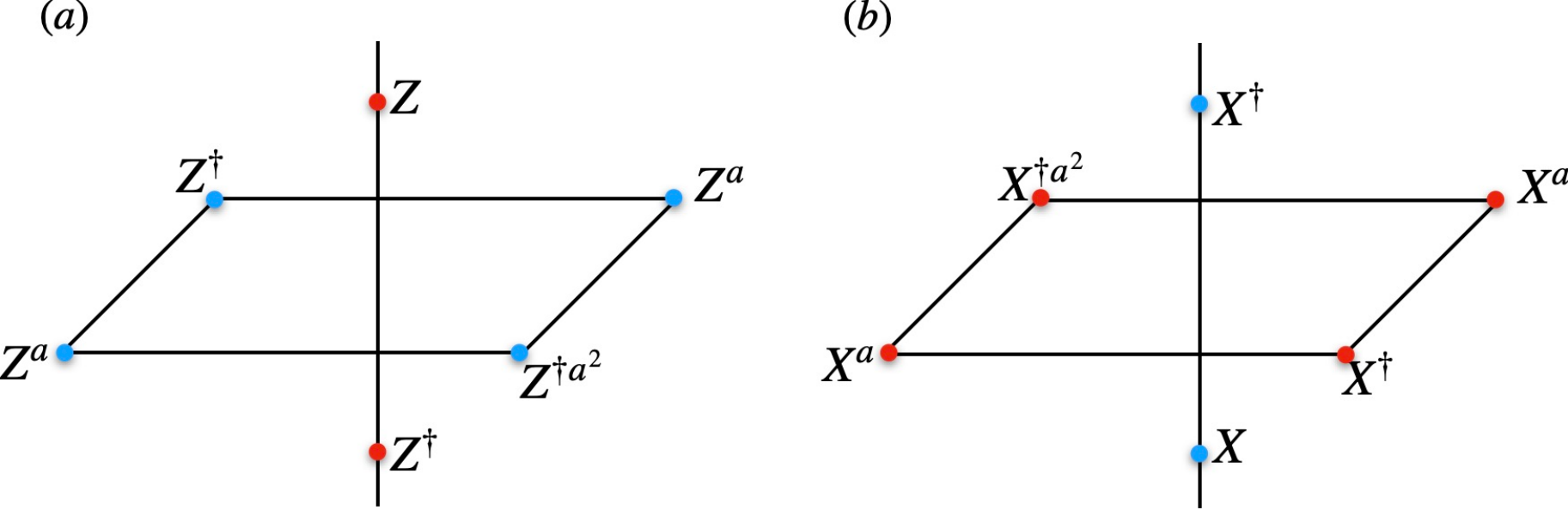}
    \caption{Exponential symmetry decoration of Planeon Lineon code: The stabilizer Hamiltonian in Eq.~\eqref{model1} is defined on qudits residing at the vertices of the $x$–$y$ plaquettes and along the $z$-links.}
    \label{planeon}
\end{figure}

To elucidate the gauge structure, we gauge the symmetry in Eq.~\eqref{3dcharges} by
coupling the boson field with a gauge potential $A_{xy},A_z$ as illustrated in Fig.~\ref{planeon}(a). 
\begin{align} 
b_r (b_{r+e_x}^{\dagger})^a (b_{r+e_y}^{\dagger})^a b_{r+e_y+e_x}^{a^2}e^{i A_{xy}}+b_r b_{r+e_z}^{\dagger}e^{i A_{z}}
\end{align}
As they mediate the interaction among three bosons $b$, the gauge fields $A_{xy}$ live on the plaquette of the square at the x-y plane, while $A_{z}$ lives on the z-links. The fields are subject to the following gauge transformations:
\begin{align} 
&A_{xy} \rightarrow A_{xy}- a^2f_{r+e_x+e_y}- f_r+af_{r+e_x}+af_{r+e_y}\nonumber\\
&A_{z} \rightarrow A_{z}- f_r
\end{align}
The generator of these gauge transformations is the generalized Gauss law:
\begin{eqnarray}
q_{r}=&a^2E_{xy,r}+E_{xy,r+e_x+e_y}-aE_{xy,r+e_y}-aE_{xy,r+e_x}\nonumber\\
&+E_{z,r}-E_{z,r+e_z}
\end{eqnarray}
Here $r$ labels the position of the vertex.

We define the gauge invariant magnetic flux operator, which also lives at the vertices of the square lattice, as illustrated in Fig.~\ref{planeon}(b).
\begin{align} 
B_{r'}=&a^2A_{z,r'}+A_{z,r'-e_x-e_y}-aA_{z,r'-e_y}-aA_{z,r'-e_x}\nonumber\\
&-A_{xy,r'}+A_{xy,r'+e_z}
\label{flux2}
\end{align}
Here $r'$ labels the position of the plaquette center.
Upon performing some algebra, one finds that the charge and flux operators are dual to each other. Consequently, the exponential flux is conserved on the $xz$ and $yz$ planes.

As before, we discretize the $U(1)$ gauge theory to a $\mathbb{Z}_N$ gauge theory. The resulting theory can be expressed as a CSS-type code, as shown in Fig.~\ref{planeon}:
\begin{eqnarray}
    H=-\sum_r\mathcal{Q}_r - \sum_r \mathcal{B}_{r'}+\text{h.c.} 
\end{eqnarray}
with $\mathcal{Q}_r$ and $\mathcal{B}_{r'}$ being the $\mathbb{Z}_N$ charge and flux operators living at the site or the cube center:
\begin{eqnarray}
\mathcal Q_r&=&Z^{-1}_{r} Z^{a}_{r+e_x} Z^{a}_{r+e_y} Z^{-a^2}_{r+e_x+e_y} Z_{r+e_z+\frac{e_x+e_y}{2}} Z^{-1}_{r-e_z+\frac{e_x+e_y}{2}} \nonumber\\
 \mathcal B_{r'}&=&X^{-a^2}_{r'} Z^{a}_{r'+e_x} X^{a}_{r'+e_y} X^{-1}_{r'+e_x+e_y} X^{-1}_{r'+e_z+\frac{e_x+e_y}{2}} X_{r'-e_z+\frac{e_x+e_y}{2}}\nonumber\\
\label{3dmodel1}
\end{eqnarray}
The model exhibits subextensive ground-state degeneracy, which also oscillate with system size:
\begin{align}
    \dim \mathcal{H}_0 = \left[\frac{\gcd(N_a,a^{L_x}-1)^{L_y}\gcd(N_a,a^{L_y}-1)^{L_x}}{\gcd(N_a,a^{L_x}-1,a^{L_y}-1)}\right]^2\label{gsd3d}
\end{align}

\subsection{Exponential symmetry decoration of X-cube}

We now consider the exponential-symmetry decoration of the 3D X-cube model~\cite{vijay2016fracton}. To set the stage, we begin with a bosonic theory defined on a cubic lattice in three dimensions illustrated in Fig.~\ref{xcube1}:
\begin{widetext}
\begin{align} 
H = -\sum_r\left[b_r (b_{r+e_x}^{\dagger})^a (b_{r+e_y}^{\dagger})^a b_{r+e_y+e_x}^{a^2}+b_r (b_{r+e_z}^{\dagger})^a (b_{r+e_y}^{\dagger})^a b_{r+e_y+e_z}^{a^2}+b_r (b_{r+e_x}^{\dagger})^a (b_{r+e_z}^{\dagger})^a b_{r+e_z+e_x}^{a^2}+\text{h.c.}\right]
\label{3dco}
\end{align}
\begin{figure}[h]
    \centering    \includegraphics[width=0.7\textwidth]{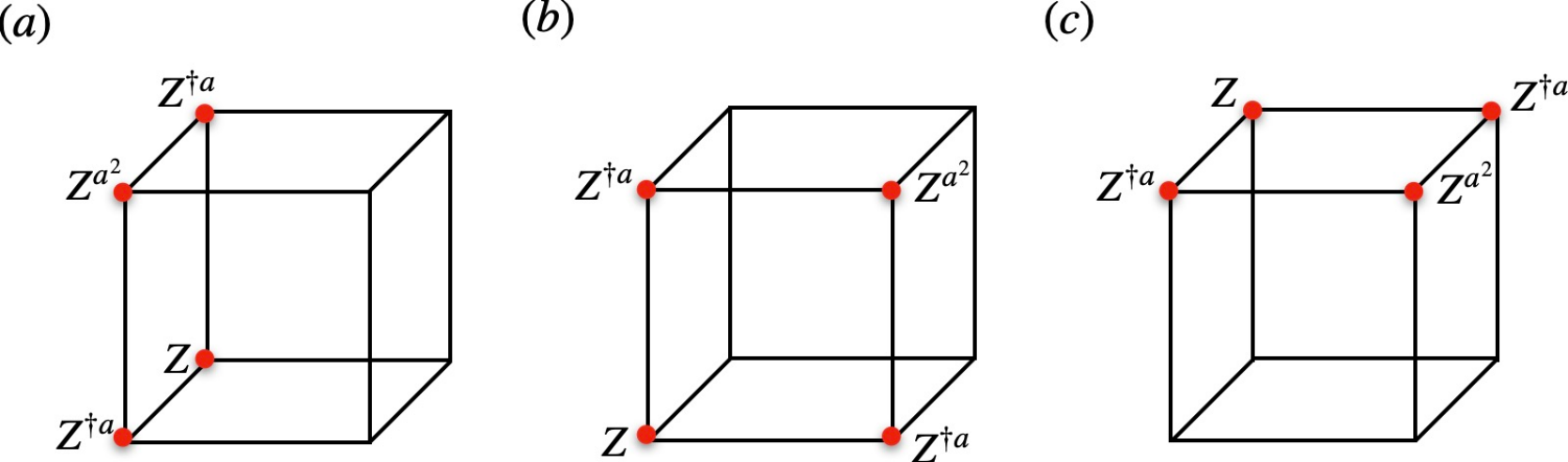}
    \caption{The 2D boson model in Eq.~\eqref{co} with ring-exchange interactions on all plaquettes. Red dots indicate bosons residing at the vertices of the cubic lattice.}
    \label{xcube1}
\end{figure}
\begin{figure}[h]
    \centering
\includegraphics[width=0.8\textwidth]{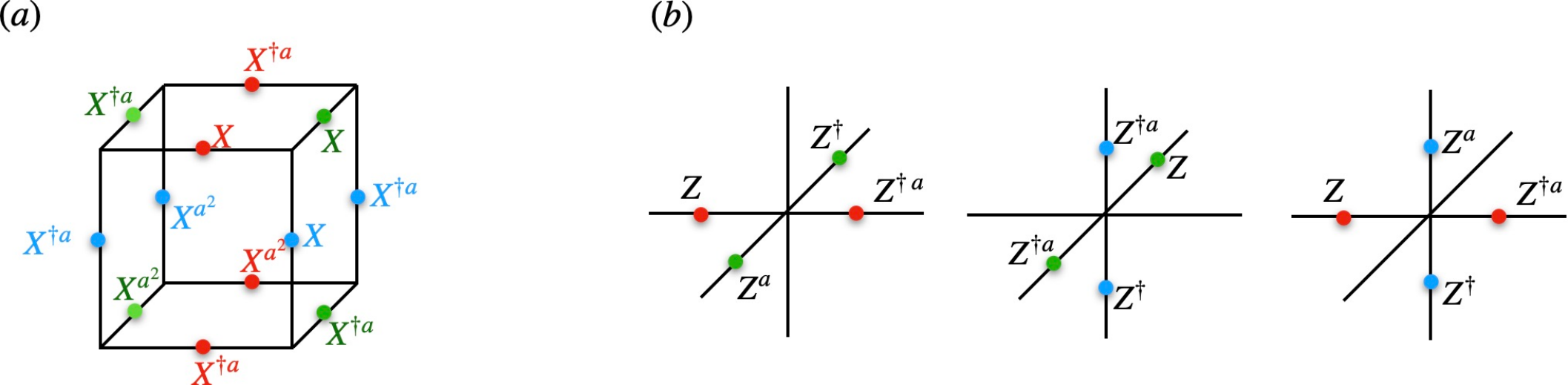}
    \caption{The exponential-symmetry–decorated X-cube model defined in Eq.~\eqref{expcube}. The figure shows the dual lattice, with qubits residing on the edges of each cube.}
        \label{xcube2}
\end{figure}

Here, $b^{\dagger}_r$ and $b_r$ represent the boson creation and annihilation operators at position $r$ in the second-quantization language. The model only contains plaquette coupling facilitates interaction among bosons at the four corners of each square. It is reminiscent of the plaquette ring-exchange model if we choose $a=1$ \cite{xu2007bond,xu2008resonating}. It is worth noting that the hopping term in Eq.~\eqref{3dco} does not preserve global charge conservation, but it does respect the subsystem exponential charge conservation law:
\begin{align} 
\sum_{r \in \text{xy-plane}} a^{x+y} q_r=G_{xy}(z),\quad
\sum_{r \in \text{yz-plane}} a^{z+y} q_r=G_{zy}(x),\quad
\sum_{r \in \text{xz-plane}} a^{x+z} q_r=G_{xz}(y)
\label{3dcon}
\end{align}
\end{widetext}
Here the subscript of $q_r$ indicates a charge located at site $r$. $G_{zy}(x)$ denotes the subsystem symmetry that measures the exponential charge on the $yz$-plane at fixed $x$.
The exponential charges on each $i-j$ plane are conserved, we refer to this as \textit{exponential subsystem} symmetry in the following discussion.
Consequently, we can view this theory as a scalar charge theory defined on an infinite-length super-lattice, where each particle is located at position $(x,y,z)$ and carries a site-dependent charge. As a result, the charge sector undergoes nontrivial transformations under translation.

To analyze the charge pattern from a generalized electromagnetism perspective, we introduce a three-component \textit{electric field}, denoted by $E_{xy,\ell}$, $E_{xz,\ell}$, and $E_{yz,\ell}$, each residing at the center of an $i-j$ plaquette. By inserting a link at the center of each plaquette where the electric field resides, we obtain a dual cubic lattice in which $E_{xy,\ell}$, $E_{xz,\ell}$, and $E_{yz,\ell}$ live on the edges of the cube, as illustrated in Fig.~\ref{xcube2}(a). These electric fields define the local charge $q_r$ as,
\begin{eqnarray}
q_r=&a^2E_{xy,r}+E_{xy,r+e_x+e_y}-aE_{xy,r+e_y}-aE_{xy,r+e_x}\nonumber\\
&+a^2E_{xz,r}+E_{xz,r+e_x+e_z}-aE_{xz,r+e_z}-aE_{xz,r+e_x}\nonumber\\
&+a^2E_{yz,r}+E_{yz,r+e_z+e_y}-aE_{yz,r+e_y}-aE_{yz,r+e_z}\nonumber\\
\label{3dexp:gua}
\end{eqnarray} 
Such charge exactly reproduce the subsystem exponential conservation law defined in Eq. \eqref{3dcon}.
Eq.~\eqref{3dexp:gua} can be viewed as a generalized `Gauss-law' that incorporates exponential subsystem charge conservation, subject to appropriate boundary conditions.

To elucidate the gauge structure associated with the exponential symmetry described in Eq. \eqref{3dcon}, we introduce a coupling between the charged boson field and a gauge potential, denoted by $A_{ij}$ living at the center of each plaquette at $i-j$ plane. 
\begin{align} 
&b_r (b_{r+e_x}^{\dagger})^a e^{iA_{xy}}(b_{r+e_y}^{\dagger})^a b_{r+e_y+e_x}^{a^2}\nonumber\\
&+b_r (b_{r+e_z}^{\dagger})^a e^{iA_{yz}}(b_{r+e_y}^{\dagger})^a b_{r+e_y+e_z}^{a^2}\nonumber\\
&+b_r (b_{r+e_x}^{\dagger})^a e^{iA_{xz}}(b_{r+e_z}^{\dagger})^a b_{r+e_z+e_x}^{a^2}
\end{align}
The gauge potential serves as the conjugate partner of the electric field and resides in the center of the plaquette of the cubic lattice. The potential is subject to gauge transformations of the form,
\begin{align} 
&A_{ij} \rightarrow A_{ij}- a^2f_{r+e_i+e_j}- f_r+af_{r+e_i}+af_{r+e_j}\nonumber\\
\end{align}

We hereby define the leading order gauge-invariant operator as the magnetic flux operator, which lives at the center of the vertex on the dual lattice $\tilde{r}$ illustrated as Fig.~\ref{xcube2}(b).
\begin{eqnarray} 
&B^x_{\tilde{r}}
=aA_{xz,\tilde{r}+\frac{e_y}{2}}-A_{xz,\tilde{r}-\frac{e_y}{2}} - aA_{xy,\tilde{r}+\frac{e_z}{2}}+ A_{xy,\tilde{r}-\frac{e_z}{2}}\nonumber\\
&B^y_{\tilde{r}}
=-aA_{yz,\tilde{r}+\frac{e_x}{2}}+A_{yz,\tilde{r}-\frac{e_x}{2}} + aA_{xy,\tilde{r}+\frac{e_z}{2}}- A_{xy,\tilde{r}-\frac{e_z}{2}}\nonumber\\
&B^z_{\tilde{r}}
= aA_{xz,\tilde{r}+\frac{e_y}{2}}-A_{xz,\tilde{r}-\frac{e_y}{2}} - aA_{yz,\tilde{r}+\frac{e_x}{2}}+ A_{yz,\tilde{r}-\frac{e_x}{2}}\nonumber\\
&B^x_{\tilde{r}}+B^y_{\tilde{r}}-B^z_{\tilde{r}}=0
\label{3dexp:flux}
\end{eqnarray}
There are three types of flux operators at each vertex but only two of them are independent.
In addition, the flux operator is subject to the conservation law,
\begin{eqnarray} 
\sum_{\tilde{r} \in \text{xy-plane}} a^{x+y} B^z_{\tilde{r}}&=&S_{xy}(z),~\nonumber\\
\sum_{\tilde{r} \in \text{yz-plane}} a^{z+y} B^x_{\tilde{r}}&=&S_{yz}(x),~\nonumber\\
\sum_{\tilde{r} \in \text{xz-plane}} a^{x+z} B^y_{\tilde{r}}&=&S_{xz}(y),
\label{flux}
\end{eqnarray}
so the exponential flux is conserved on the 2D plane.

After discretizing the $U(1)$ gauge theory to $\mathbb{Z}_N$, the resulting gauge theory can be expressed in terms of a CSS-type code as Fig.~\ref{xcube2},
\begin{eqnarray}\label{expcube}
    H=-\sum_r\mathcal{Q}_r - \sum_{\tilde{r}} \sum_{a=x,y,z}\mathcal{B}^a_{\tilde{r}}+\text{h.c.} 
\end{eqnarray}
with $\mathcal{Q}_r$ and $\mathcal{B}^a_{\tilde{r}}$ manifest the local charge and flux operators, defined in the center of the cube and the vertex
as illustrated in Fig.~\ref{xcube2}. This model can be viewed as an exponential decoration of the X-cube model proposed in Ref.~\cite{vijay2016fracton}. A detailed discussion, including its ground-state degeneracy and Wilson operators, is provided in Appendix \ref{app:expcube}.

Similar to the 2d examples, the fracton topological order in Eq.~\eqref{expcube} only exist when $a \neq 0 $ mod rad ($N$), so the symmetries we begin with are global symmetries. If $a = p \,\text{rad}(N)$, for some $p\in \mathbb Z$, one can infer that there exists a finite integer $m$ such that $a^{m}=0$ mod $N$. Consequently, the exponential symmetry only acts on a finite number of degrees of freedom for sites $|r|<m$, and the exponential symmetry operator becomes a local symmetry in the thermodynamic limit.

\section{Discussion and Outlook}

In this work, we present a class of $\mathbb{Z}_N$ gauge theories obtained by gauging exponential polynomial symmetries. In two dimensions, these theories, formulated through stabilizer Hamiltonians, share key features with conventional topological orders: they possess topological ground-state degeneracy, support anyonic excitations created by non-local operators, and exhibit long-range entanglement. At the same time, they differ in important ways: their charge and flux excitations have restricted mobility on the lattice, and their ground-state degeneracy depends on parameters set at the UV scale. We further generalize the concept of exponential symmetry to three dimensions, introducing a broader class of fracton models. The stabilizer codes and discrete gauge theories developed here open potential new avenues for exploring generalized symmetries and fracton gauge theories. We conclude by outlining several open questions for future investigation:


1) While the majority of topological phase transitions between deconfined gauge theories and Higgsed (or confined) phases have been explored in the early literature, it would be compelling to investigate possible phase transitions by adding onsite transverse fields that could potentially trigger a transition between the deconfined phase displayed in this manuscript and a confined phase.

2) Inspired by the double-semion model as a twisted $\mathbb{Z}_2$ gauge theory in 2D, we anticipate that one could develop a similar twisted gauge theory with exponential charge conservation. Another relevant direction would be to extend the $\mathbb{Z}_N$ gauge theory explored in this manuscript into U(1) and seek possible deconfined U(1) phases by adding Chern-Simons-like couplings.

3) As highlighted in Sec.~\ref{sec:one}, exponential symmetries generators do not commute with spatial translations. As a result, spatial symmetry defects, such as dislocations, can permute between different topological charge sectors, making it engrossing to investigate the effects of lattice defects as they could potentially trap topological zero modes.


 \acknowledgments 
 YY acknowledges support from NSF under award number DMR-2439118. This work is also supported by the DOE Grant No. DE-FG02-06ER46316 (G. D. and C.C.).

\appendix
\label{appendix}

\section{Ground State Degeneracy and Global Constraints in Exponential Charge Gauge Theory}\label{global_constraints}
In this appendix, we derive the ground state degeneracy in Eq. \eqref{gsd} for the exponential charge model in Eq. \eqref{model1}.
First, let us note there are two $\mathbb{Z}_N$ degrees of freedom per site, giving rise to a total Hilbert space of dimension $\dim \mathcal H = N^{2N_{\text{sites}}}$, where $N_{\text{sites}}$ is the number of sites on the lattice. There are, as well, two $\mathbb{Z}_N$ operators  with $N$ distinct eigenvalues $\exp{2\pi q_r\ N }$ - the charge $\mathcal{Q}_r$ and the flux $\mathcal{B}_r$ - per site, being able to distinct and label $N^{2N_{\text{sites}}}$ states. The degeneracy, however, comes from the fact that under periodic boundary conditions, not all the eigenvalues $q_r=0, \ldots, N-1$ for different lattice sites are independent. Instead, they are subject to  global constraints, as in Eq. \eqref{conserv}, which here we carefully derive.


The $\rho$ factors are extremely important, telling us how many  distinct states are actually labeled by the same $\mathcal{Q}_r$ and $\mathcal{B}_r$ eigenvalues. We note that the flux $\mathcal{B}_r$ operators also obey the same equations as $\mathcal{Q}_r$ in Eq. \eqref{conserv} and that each one of the constraints contributes to the ground state degeneracy. The role of each $\rho$ factor is to ``reduce'' the $\mathbb{Z}_N$ operators in the products \eqref{conserv} down to a $\mathbb{Z}_{N_a/\rho}$ ones. The ground state degeneracy is given by the number of such different configurations $\left(N\times N_a/\rho_1\times N_a/\rho_2\times N_a/\rho_{12}\right)^2$, simplifying to
\begin{eqnarray}
    \dim \mathcal{H}_0 = \left[N \gcd(N_a,a^{L_x}-1)\gcd(N_a,a^{L_y}-1)\right.\nonumber\\
    \left.\gcd(N_a,a^{L_x}-1,a^{L_y}-1)\right]^2,
\end{eqnarray}
where the power of 2 takes into account the constraints for both the charge and flux operators.

The way we count the global constraints in Eq. \eqref{conserv} is to find the sets of solutions $t_r$ mod $N$ to the product of charge operators in different sites $r = (x,y)$
\begin{eqnarray}\label{prod}
    \prod_r \mathcal{Q}_r^{t_r} = \mathds{1}.
\end{eqnarray}
We inherit the discussion present in Ref.~\cite{Watanabe2022} based on Euler's totient theorem; we consider the case when $a$ and $N$ are not necessarily coprime but $a\neq 0$ mod rad($N$). In such a case, one can always define $N_a$ as the greatest divider of $N$ that is coprime to $a$. From Euler's theorem, there always exists a finite integer $\varphi (N_a)$ such that $a^{\varphi (N_a)}-1=0 \,(\text{mod } N_a)$ with $\varphi (n)$ being the Euler's totient function, allowing us to find the $\rho$ factors.

In the following we show that the possible $t_r$ solutions are $t_r=t$, $\rho_1 a^{x}$, $\rho_2 a^{y}$, $\rho_{12} a^{x+y}$, for integers $t, \rho_1, \rho_2, \rho_{12}$. The functional form of these solutions can be traced back to the conservation laws of usual charge and exponential charges in Eq. \eqref{charges}. To see this, we use the explicit form of $\mathcal{Q}_r$ and rewrite Eq. \eqref{prod} in terms of Pauli operators coming from different $A_r$ acting on a single site $r$
\begin{eqnarray}
    \prod_r \left(X_r\right)^{(a+1)t_r-at_{r-e_y}-t_{r+e_y}} \left(\bar{X}_r\right)^{(a+1)t_r-t_{r+e_x}-at_{r-e_x}} = \mathds{1}.\nonumber
\end{eqnarray}
We thus see that the only way for the constraint \eqref{prod} to hold globally is that each term in the product above reduces to the identity. The problem then translates into solving a pair of recursive modular equations
\begin{eqnarray}
    (a+1)t_r-at_{r-e_y}-t_{r+e_y} &=& 0\quad \text{mod }N,\nonumber\\
    (a+1)t_r-t_{r+e_x}-at_{r-e_x} =&& 0\quad \text{mod }N,\label{solve_eq}
\end{eqnarray}
which we solve, in some detail. The ground state degeneracy, in the end, boils down to counting how many independent solutions such equations admit.

First, let us note that the constant solution $t_r=t$, with $t$ an integer mod $N$ satisfies the two equations above, providing us with $N$ independent solutions. This is the $N$ factor in the GSD expression in Eq. \eqref{gsd}. Another class of solutions, which is less trivial than the constant one is the exponential
\begin{eqnarray}\label{expx}
    t_r = \rho_1\, a^x \quad \text{mod}\, N
\end{eqnarray}
where $x$ is the horizontal coordinates for the lattice point $r$. As mentioned in the main text, we consider only $a\neq 0$ mod rad($N$) such that $a^x$ never vanishes mod $N$. Under periodic boundary conditions, the values $\rho_1$ must obey
\begin{eqnarray}
    \rho_1 a^x = \rho_1a^{x+L_x}\quad \text{mod }N  \quad \forall\,  x=1, \ldots, L_x, \nonumber\\
    \Rightarrow \rho_1\left(a^{L_x}-1\right) = 0\quad
 \text{mod}\,\, N_a
\end{eqnarray}
whose solutions are
\begin{eqnarray}
    \rho_1= \dfrac{N_a}{\gcd(N_a,a^{L_x}-1)}k,
\end{eqnarray}
where $k$ is an integer number mod $\gcd(N_a,a^{L_x}-1)$. Thus, the ansatz in Eq. \eqref{expx} assumes $ \gcd(N_a,a^{L_x}-1)$ distinct solutions, contributing with this number to the ground state degeneracy expression in Eq. \eqref{gsd}. From similar arguments, the class of solutions
\begin{eqnarray}\label{expy}
    t_r = \rho_2\, a^y \quad \text{mod}\, N
\end{eqnarray}
contributes with a factor of $\gcd(N_a,a^{L_y}-1)$ to GSD. Finally, 
\begin{eqnarray}
    t_r = \rho_{12} \,a^{x+y},
\end{eqnarray}
are also solutions to the recursive equations \eqref{solve_eq}. Here, the periodic boundary conditions require simultaneously that
\begin{eqnarray}
    \rho_{12} \left(a^{L_x}-1\right) = 0\, \text{mod}\,\, N_a,\nonumber\\
    \text{and}\quad \rho_{12}  \left(a^{L_y}-1\right) = 0\, \text{mod}\,\, N_a,
\end{eqnarray}
which can be solved by
\begin{eqnarray}
    \rho_{12}  = \dfrac{N_a}{\gcd(N_a, a^{L_x}-1, a^{L_y}-1)} p,
\end{eqnarray}
with $p$ an integer mod $\gcd(N_a, a^{L_x}-1, a^{L_y}-1)$. The integer $p$ parameterizes how many independent solutions the ansatz in Eq. \eqref{expy} assumes, contributing with a factor of $\gcd(N_a, a^{L_x}-1, a^{L_y}-1)$ to the expression for the GSD.

Using similar arguments, we can also show that the flux operators $\mathcal B_r$ are subject to global constraints
\begin{eqnarray}
    \prod_r \mathcal{B}_r^{\tilde t_r} = \mathds{1},
\end{eqnarray}
as well.
From  inspection, it is straightforward to show that $\tilde t_r$ obeys similar equations as $t_r$ in \eqref{solve_eq} and, consequently, obeys similar solutions as well. When taking these into account,  the ground state degeneracy in Eq. \eqref{gsd} acquires an overall power of 2.

\subsection{Wilson Lines in Exponential Charge Gauge Theory}\label{app_wilson}

We now apply this protocol to our model, proposed in Sec.~\ref{sec:two}. Upon Higgsing, we used an exponential map from $A_{i, r}$ and $E_{i,r}$ to the $\mathbb{Z}_N$ operators $Z$ and $X$.
The ground state is projected to zero local fluxes with eigenvalue $\mathcal{B}_r=1$ for all lattice sites $r$. Thus, the total flux on any open or closed area is subject to $\prod_r \mathcal{B}_r=  1$. In addition, one can also derive that the eigenvalues of the exponential fluxes also multiply up to one, $\mathbb{Z}_N$ analogs of the fluxes in Eq. \eqref{con2}.

Consider now the ground state wave function on an open cylinder, as shown in Fig.~\ref{flux_area}. We focus on the Wilson lines that are defined along the y-loops, associated to the first two conservation laws in Eq.~\eqref{con2} yielding,
\begin{widetext}
\begin{align}
&\prod_{r\in \cal A} \mathcal{B}_r = \prod_{i}  Z^a(0,i)\, Z^\dagger(1,i) \prod_{j} Z^{\dagger a}(x_0-1,j)\, Z(x_0,j),
 \nonumber\\
&\prod_{r\in\cal A}  \mathcal{B}_r^{a^{x_0-x}}  = \prod_{i}  Z^{a^{x_0}}(0,i)\,  Z^{-a^{x_0}}(1,i) \prod_{j}  Z^\dagger(x_0,j)\, Z(x_0,j).
\label{eq:qxqy-conserve-2}
\end{align}
Based on Eq.~\eqref{eq:qxqy-conserve-2}, the total flux on an open cylinder is reduced to the holonomy operators localized at the boundaries,
\begin{align}
W^{(1)}(x) & =  \prod_{i}  Z^a(x,i)\, Z^\dagger(x+1,i) ,  \quad \text{and}\quad W^{(2)}(x)  =  \prod_{i}  Z(x,i)\, Z^\dagger(x+1,i). 
\end{align}
\end{widetext}
It follows from the flux conservation law that these two operators are constrained to obey
\begin{align}\label{product1}
W^{(1)} (x)\,  W^{ (1)\dagger } (x+1) =1,\nonumber\\
W^{(2)\dagger a}(x-1)\, W^{(2)}(x)=1.
\end{align}

\begin{widetext}
Following the same procedure for the last two conservations laws in Eq.~\eqref{con2}, when integrated on a cylinder $\mathcal A$ as in Fig. \ref{flux_area},  they decompose into products
\begin{align}
&\prod_{r\in \cal A} \mathcal{B}_r^{a^{L_y-y}} = \prod_{i}  Z^{a^{L_y-i+1}}(0,i)\, Z^{\dagger a^{L_y-i}}(1,i) \prod_{j}  Z^{\dagger a^{L_y-j+1}}(x_0-1,j)\,Z^{a^{L_y-j}}(x_0,j),
 \nonumber\\
&\prod_{r\in \cal A} \mathcal{B}_r^{a^{x_0-x} a^{L_y-y}} = \prod_{i}  Z^{a^{L_y-i+x_0}}(0,i) \, Z^{\dagger a^{L_y-i+x_0}}(1,i) \prod_{j} Z^{\dagger a^{L_y-j}}(x_0,j)\,Z^{a^{L_y-j}}(x_0,j).
\label{eq:qxqy-conserve-3}
\end{align}
These engender another set of Wilson line operators
\begin{align}
W^{(3)}(x) & =  \prod_{i}  Z^{a^{L_y-i+1}}(x,i)\, Z^{\dagger a^{L_y-i}}(x+1,i) ,  \quad\text{and}\quad  W^{(4)}(x)  =  \prod_{i}  Z^{a^{L_y-i}}(x,i)\, Z^{\dagger a^{L_y-i}}(x+1,i). 
\end{align}
\end{widetext}
Following the flux conservation laws, these operators obey
\begin{align}
W^{(3)} (x) W^{(3)\dagger }(x+1) =1,\nonumber\\
W^{(4)\dagger a}(x-1)W^{(4)}(x)=1.\label{product2}
\end{align}
Similarly, one can define the dual operators going along the $x$-direction. We characterize the charge sectors  by integrating the conserved quantities in Eq. \eqref{charges} on an open cylinder  along the $y-$ direction (a rotated cylinder when compared to the one in Fig. \ref{flux_area}.,
\begin{eqnarray}
V^{(1)}_x(y)  &=&  \prod_{i}  X^\dagger(i,y)\, X^{a} (i,y+1), \nonumber\\ V^{(2)}_x(y)    &=&  \prod_{i}  X^\dagger(i,y) \, X(i,y+1) \nonumber\\
V^{(3)}_x(y)    &=&  \prod_{i}  X^{\dagger a^{i}}(i,y)\, X^{a^{i+1}}(i,y+1) , \nonumber\\ 
V^{(4)}_x(y)    &=&  \prod_{i}  X^{\dagger a^{i}}(i,y)\,  X^{a^{i}}(i,y+1), 
\end{eqnarray}

Likewise, the Wilson lines composed of $\bar{X},\bar{Z}$ also emerge at the boundary of cylinder regions, 
\begin{eqnarray}
W_x^{(1)}(y) & =&  \prod_{i}  \bar{Z}^{a}(i,y)\, \bar{Z}^\dagger(i,y+1) ,  \nonumber\\
W_x^{(2)}(y) &  =&  \prod_{i}  \bar{Z}(i,y)\,\bar{Z}^\dagger(i,y+1), \nonumber\\
W_x^{(3)}(y) & = & \prod_{i}  \bar{Z}^{a^{L_x-i+1}}(i,y)\,\bar{Z}^{\dagger a^{L_x-i}}(i,y+1) ,  \nonumber\\
~W_x^{(4)}(y) &   = & \prod_{i}  \bar{Z}^{a^{L_x-i}}(i,y)\,\bar{Z}^{\dagger a^{L_x-i}}(i,y+1), \nonumber\\
V^{(1)}_y(x)  &=&  \prod_{i}  \bar{X}^\dagger(x,i)\,\bar{X}^a(x+1,i) , 
\nonumber\\
V^{(2)}_y(x)  &=&  \prod_{i}  \bar{X}^\dagger(x,i)\,  \bar{X}(x+1,i),\nonumber\\
V^{(3)}_y(x)  &=&  \prod_{i}  \bar{X}^{\dagger a^{i}}(x,i)\, \bar{X}^{a^{i+1}}(x+1,i) ,  
\nonumber\\
V^{(4)}_y(x)&=&\prod_{i}  \bar{X}^{\dagger a^{i}}(x,i) \, \bar{X}^{a^{i}}(x+1,i), 
\end{eqnarray}
where all these $V^{(i)}$ and $W^{(i)}$ obey similar conditions to Eq. \eqref{product1} and \eqref{product2}.

\section{Exponential Symmetry Decoration of X-Cube}\label{app:expcube}

\subsection{Ground State Degeneracy}

The phase described by the ground state of the Hamiltonian in Fig.~\ref{xcube2} is gapped and topologically ordered. The model sits at an exactly solvable point, allowing us to solve the system exactly. We now count the ground state degeneracy when the system is put on a torus, which can be non-trivial on compact lattices. Every term in the Hamiltonian commute with each other so we can diagonalize all of them simultaneously. Here, we focus on the ground state space, corresponding to the eigenvalues $+1$, minimizing the Hamiltonian energy

\begin{equation}
    \mathcal{H}_0 = \{\ket{\psi}\in \mathcal{H}\, \big | \, \mathcal{Q}_r \ket{\psi} = +1\ket{\psi} \text{ and }\mathcal{B}^a_r \ket{\psi} = +1\ket{\psi}\}.
\end{equation}
The ground state space is characterized by the local requirement of vanishing $\mathbb{Z}_N$ charge and flux everywhere. In such case, one can distinguish elements in $\mathcal{H}_0$, if degenerate, only through global features, which is the essence of topologically ordered states. 

In the following, we assume periodic boundary conditions for the lattice on a 3-torus. This introduces consistency requirements on the field boundary configurations such that the conserved quantities in Eq. \eqref{3dcharges} are well defined. In terms of the $\mathbb{Z}_N$ degrees of freedom, the conservation equations in Eq. \eqref{3dcharges} translate into
\begin{align}\label{conserv1}
    \prod_{x,y} \mathcal{Q}_r^{\rho_1 a^{x+y}}=1, \nonumber\\
    \prod_{z,y} \mathcal{Q}_r^{\rho_2 a^{z+y}}=1,\nonumber\\
    \prod_{x,z} \mathcal{Q}_r^{\rho_3 a^{x+z}}=1,
\end{align}
and similarly for the flux conservations in terms of the $\mathcal{B}^a_r$ operators. 

\begin{align}
\label{conserv2}
    \prod_{x,y} (\mathcal{B}^z_r)^{\rho_1 a^{x+y}}=1, \nonumber\\
    \prod_{z,y} (\mathcal{B}^x_r)^{\rho_2 a^{z+y}}=1,\nonumber\\
    \prod_{x,z} (\mathcal{B}^y_r)^{\rho_3 a^{x+z}}=1,
\end{align}

In the above, $N_a$ is the greatest divider of $N$ that is coprime to $a$, and the $\rho$ factors are given by
\begin{eqnarray}
\rho_1 &=& \dfrac{N_a}{\gcd(N_a,a^{L_x}-1,a^{L_y}-1)},\nonumber\\
    \rho_2 &=& \dfrac{N_a}{\gcd(N_a,a^{L_y}-1,a^{L_z}-1)},\nonumber\\
    \rho_{12} &=& \dfrac{N_a}{\gcd(N_a,a^{L_x}-1,a^{L_z}-1)},
\end{eqnarray}
where $L_x$ and $L_y$ are the linear sizes of the lattice in the $x$ and $y$ directions. As the exponential charge is conserved on each i-j plane, we expect the ground state entropy is sub-extensive. The ground state degeneracy can be count as follow. From Eq.~\eqref{conserv1}, it is not hard to conclude that the exponential charge conservation on each i-j plane give rise to a degeneracy of $\gcd(N_a,a^{L_i}-1,a^{L_j}-1)$. Thus, it seems that counting all i-j planes engenders a a degeneracy of $\prod_{ijk}(\gcd(N_a,a^{L_i}-1,a^{L_j}-1)^{L_k}|\epsilon^{ijk}|)$. However, the charge conservation laws in Eq.~\eqref{conserv1} are not independent due to the fact that,
\begin{align}
    \prod_{z}\left(\prod_{x,y} \mathcal{Q}_r^{\rho_1 a^{x+y}}\right)a^z
    =\prod_{x}\left(\prod_{z,y} \mathcal{Q}_r^{\rho_2 a^{z+y}}\right)a^x\nonumber\\
    =\prod_{y}\left(\prod_{x,z} \mathcal{Q}_r^{\rho_3 a^{x+z}}\right)a^z
\end{align}
Thus, we need to divide the double counting by $2\gcd(N_a,a^{L_x}-1,a^{L_y}-1,a^{L_z}-1)$. Likewise, from 
Eq.~\ref{conserv2}, it is not hard to conclude that the exponential flux conservation $B^k$ on each i-j plane give rise to a degeneracy of $\gcd(N_a,a^{L_i}-1,a^{L_j}-1)$. Thus, it seems that counting flux on all i-j planes engenders a a degeneracy of $\prod_{ijk}(\gcd(N_a,a^{L_i}-1,a^{L_j}-1)^{L_k}|\epsilon^{ijk}|)$. However, the subsystem flux conservation laws in Eq.~\ref{conserv2} are not independent due to the fact that,
\begin{align}
    \prod_{z}\left(\prod_{x,y} (\mathcal{B}^z)_r^{\rho_1 a^{x+y}}\right)a^z
    \prod_{x}\left(\prod_{z,y} (\mathcal{B}^x)_r^{\rho_2 a^{z+y}}\right)a^x\nonumber\\
    \prod_{y}\left(\prod_{x,z} (\mathcal{B}^y)_r^{\rho_3 a^{x+z}}\right)a^z=1
\end{align}
Indeed, we need to divide the double counting factor by $\gcd(N_a,a^{L_x}-1,a^{L_y}-1,a^{L_z}-1)$. This concludes the ground state entropy is:
\begin{eqnarray}
    S_{GS}= 2L_z \ln(\gcd(N_a,a^{L_x}-1,a^{L_y}-1))\nonumber\\+2L_x \ln(\gcd(N_a,a^{L_x}-1,a^{L_z}-1))\nonumber\\
+2L_y \ln(\gcd(N_a,a^{L_x}-1,a^{L_z}-1))\nonumber\\
-3\ln(\gcd(N_a,a^{L_x}-1,a^{L_z}-1,a^{L_y}-1).\label{gsd2}
\end{eqnarray}

This result exhibits a subextensive functional dependence that oscillates polynomially with system size, resembling certain phenomena in 3D fracton physics~\cite{Chamon2005-fc,Haah2011-ny,vijay2016fracton,shirley2018fractional,prem2017emergent,oh22a,oh22b}.

\subsection{Wilson operators}
Topological quantum field theories (TQFTs) are characterized by their ground-state degeneracy on closed manifolds, the holonomies generated by Wilson line operators, and anyonic excitations with nontrivial braiding. Similar characterization applies to many 3D fracton stabilizer codes—realized in higher-rank gauge theories beyond the TQFT paradigm—though their ground-state degeneracy depends on ultraviolet (UV) conditions, and their excitations exhibit restricted, subdimensional mobility.
From Eq.~\eqref{expcube}, our goal is to identify Wilson operators that uniquely determine the holonomies and govern quasiparticle dynamics. Following Ref.~\cite{wen03}, we construct these operators by examining nonlocal, gauge-invariant operators under gauge transformations. We begin by identifying open Wilson string operators that commute with the stabilizer except at their endpoints, where the flux excitation $B^a$ resides.
\begin{align} 
&F_1=\prod^s_{i}  X^{a^i}_{r-ie_x},\nonumber\\
&F_2=\prod^s_{i}  X^{a^i}_{r-ie_y},\nonumber\\
&F_3=\prod^s_{i}  X^{a^i}_{r-ie_z},\nonumber\\
\label{wil1}
\end{align}
As in the well-known X-cube model, the flux excitation is restricted to move along a single spatial dimension, so the corresponding Wilson line operator cannot bend.
Another Wilson operator can be readily identified—this one taking the form of an open membrane.
\begin{align} 
&G_1=\prod^s_{i,j} (Z_{r+ie_x+je_y})^{a^{i+j}},\nonumber\\
&G_2=\prod^s_{i,j} (Z_{r+ie_y+je_z})^{a^{i+j}},\nonumber\\
&G_3=\prod^s_{i,j} (Z_{r+ie_z+je_x})^{a^{i+j}},\nonumber\\
\label{wil2}
\end{align}
The operators $G_i$ create charge-type excitations at the membrane’s corners. These membranes can be extended or deformed, with the excitations carried by their corners.
After identifying these topological excitations, we examine the holonomy sectors and ground-state degeneracy. Extending a Wilson operator along a noncontractible loop annihilates the excitation at the string end (or membrane corner). Consequently, Wilson operators defined along noncontractible regions determine the topological sectors and holonomies of the degenerate ground state, which can be interpreted as winding numbers of electric or magnetic fields.

Although Eq.~\eqref{wil1} appears to define an extensive set of Wilson operators, the ground-state manifold is fixed by the number of independent ones. We start with the flux operators in Eq.~\eqref{wil1}. To count the independent string operators $F_1$ oriented along $x$, we pair adjacent strings in the $yz$ plane:
\begin{align} 
&F^z_1(y,z)=F_1(y,z)F^{-a}_1(y,z-1)=\prod^s_{i}  X^{a^i}_{r-ie_x} (X^{a^i}_{r-ie_x-e_z})^{-a},\nonumber\\
&F^y_1(y,z)=F_1(y,z)F^{-a}_1(y-1,z)=\prod^s_{i}  X^{a^i}_{r-ie_x} (X^{a^i}_{r-ie_x-e_y})^{-a},\nonumber\\
\end{align}
$F^z_1$ contains two parallel $F_1$ strings that are separated by lattice vector $e_z$. From the charge conservation law, we can conclude that $F^z_1(y,z)(F^z_1(y-1,z))^{-a}=1$. This infers that as long as we fixed $F^z_1(0,z)$, the other values of $F^z_1(y,z)$ should be fixed. In addition, under periodic boundary conditions, we also have $F^z_1(0,z)(F^z_1(L_y,z))^{-a^{L_y}}=(F^z_1(0,z))^{1-a^{L_y}}=1$, this restricts the value of $F^z_1(0,z)$ to into a $Z_{\gcd(a^{L_y}-1,a^{L_x}-1,N)}$ number. Consequently, $F^z_1(0,z)$ can take $Z_{\gcd(a^{L_y}-1,a^{L_x}-1,N)}$ values for all choice of $z$. This generates a ground state entropy of $L_z \ln(Z_{\gcd(a^{L_y}-1,a^{L_x}-1,N)}))$. Likewise, one can also conclude that $F^y_1(y,0)$ take $Z_{\gcd(a^{L_y}-1,a^{L_x}-1,N)}$ values for all choice of $y$ and generates additional ground state entropy of $L_y \ln(Z_{\gcd(a^{L_z}-1,a^{L_x}-1,N)}))$. However, we need to be careful with the double counting since the choice of $F^y_1(0,0),F^y_1(1,0),F^z_1(0,0),F^z_1(0,1)$ are not independent. After excluding the double counting in entropy, we figured out that the entropy generated by the topological sector of $F_1$ is:
\begin{eqnarray}
  &L_z \ln(Z_{\gcd(a^{L_y}-1,a^{L_x}-1,N)}))+ L_y \ln(Z_{\gcd(a^{L_z}-1,a^{L_x}-1,N)}))\nonumber\\
  &- \ln(Z_{\gcd(a^{L_z}-1,a^{L_x}-1,a^{L_z}-1,N)})) \nonumber
\end{eqnarray}
Following the same spirit, one can count the entropy generated by the topological sector of $F_2, F_3$ and the total entropy of the topological flux sector is 
\begin{widetext}
\begin{eqnarray}
  &2L_z \ln(Z_{\gcd(a^{L_y}-1,a^{L_x}-1,N)}))+ 2L_y \ln(Z_{\gcd(a^{L_z}-1,a^{L_x}-1,N)}))\nonumber 2L_x \ln(Z_{\gcd(a^{L_z}-1,a^{L_y}-1,N)}))- 3\ln(Z_{\gcd(a^{L_z}-1,a^{L_x}-1,a^{L_z}-1,N)})) 
\end{eqnarray}
\end{widetext}

\section{Exponential Symmetry from Symmetry Enriched Toric Code}\label{sec:appc}

As we discussed in Sec.~\ref{sec:two}, the exponential gauge theory can be viewed as a translation symmetry-enriched topological order in which lattice translations permute (or transmute) distinct gauge charges. A related question is whether the exponential gauge theory can be obtained directly by starting from a conventional symmetry-enriched topological (SET) phase and then gauging its global symmetry.
More broadly, the connection between fracton gauge theories, whose quasiparticles exhibit mobility constraints, and symmetry-enriched topological order has been examined in Ref.~\cite{williamson2019fractonic}. In that framework, one considers the toric code enriched by a global symmetry $G$. Moving an anyon along a spatial path changes the total $G$-charge carried by the system; equivalently, the string operators that transport anyons are charged under $G$. Upon gauging $G$, the resulting gauge theory can display fractonic features, in the sense that certain excitations acquire restricted mobility.

Motivated by this idea, we explore an alternative route to exponential gauge theory by starting from the usual toric code, in which the charge excitation is enriched by a modulated symmetry $G$. Upon gauging $G$, the resulting theory becomes reminiscent of the exponential gauge theory studied in Sec.~\ref{sec:two}.
We begin with the standard $\mathbb{Z}_N$ gauge theory, realized as the toric code on a square lattice. 
\begin{equation}
{H}=-\sum_{ r} (B_{ r}+B_{ r}^\dag) - \sum_r (A_r+A_r^\dag),
	\label{}
\end{equation}
whose the plaquette terms $B_{{\tilde r}}$ and vertex terms $A_r$ are defined as
\begin{align}
B_{ r}=\begin{tikzpicture}[scale=0.50, baseline={([yshift=-.5ex]current  bounding  box.center)}]
\draw (1,1) -- (-1,1);
\draw (-1,1) -- (-1,-1);
\draw (-1,-1) -- (1,-1);
\draw (1,-1) -- (1,1);
\draw (0,1.4) node {$Z^\dag$};
\draw (0,-1.4) node {$Z$};
\draw (-1.45,0) node {$Z^\dag$};
\draw (1.45,0) node {$Z$};
\draw (-1.35,-1.25) node {$ r$};
\end{tikzpicture}\, ,
&&
A_r = \begin{tikzpicture}[scale=0.50, baseline={([yshift=-.5ex]current  bounding  box.center)}]
\draw (0,-1) -- (0,1);
\draw (-1,0) -- (1,0);
\draw (0,1.4) node {$X$};
\draw (0,-1.4) node {$X^\dag$};
\draw (-1.5,0) node {$X^\dag$};
\draw (1.4,0) node {$X$};
\draw (0.3,0.35) node {$r$};
\end{tikzpicture}\, .
\end{align}
The charge-$e$ excitations are created and transported by acting with a string of $Z$ operators along a path, via the string operator
\begin{equation}
	T_{x}=\prod_{0\leq x< m}Z_{(x,y),\hat{x}}^\dag 
	\, ,
	\label{eq:W}
\end{equation}
creates an $e$-$\bar e$ pair at the endpoints $(0,y)$ and $(m,y)$. Here the subscript ${\vr,\hat{\imath}}$ labels the edge emanating from the site $r=(x,y)$ in the $\hat{\imath}$ direction. Equivalently, $T_x$ is a string operator that translates an $e$ charge by $m$ sites along $\hat{x}$. 

Likewise, one can define a vertical string operator $T_y$ that moves the charge by $m$ sites along $\hat{y}$.
\begin{equation}
	T_{y}=\prod_{0\leq y< m}Z_{(x,y),\hat{y}}^\dag 
\end{equation}

Now we consider two types of exponential symmetries $\Z_N$ symmetries, generated by the following operators,
\begin{align}
  S_{\hat{x}} &= \prod_r X^{f(x)}_{r,\hat{x}}\,,~~
  S_{\hat{y}} = \prod_r X^{f(y)}_{r,\hat{y}}\,, \nonumber\\
  & f(y)=a^y-a^{y-1}\ \text{for} ~y>0,\qquad f(0)=1
  \label{eqn:SinZN}
\end{align}
The toric code Hamiltonian respects both symmetries. The $e$–$\bar e$ pair created by the $T$ string acting on $\ket{0}$ carries a nonzero $S_{\hat{x}}$ quantum number:
\begin{equation}
	S_{\hat{x}} T_x\ket{0} = e^{\frac{2\pi i}{N}a^m}T_x\ket{0}, ~S_{\hat{y}} T_y\ket{0} = e^{\frac{2\pi i}{N}a^m}T_y\ket{0}
\end{equation}
In essence, $S_{\hat{x}}$ defines and measures the $\Z_N$ exponential symmetry charge carried by an $e$ excitation along $\hat{x}$. Consequently, translating an $e$ anyon by $m$ sites along $\hat{x}$ multiplies the state’s $S_{\hat{x}}$ eigenvalue by $a^{m}$; if the $S_{\hat{x}}$ symmetry is to be preserved, this motion is allowed only when $a^{m}\equiv 1 \ (\mathrm{mod}\ N)$. An analogous statement holds for $S_{\hat{y}}$. 

\begin{figure}[h]
    \centering
   \includegraphics[width=0.46\textwidth]{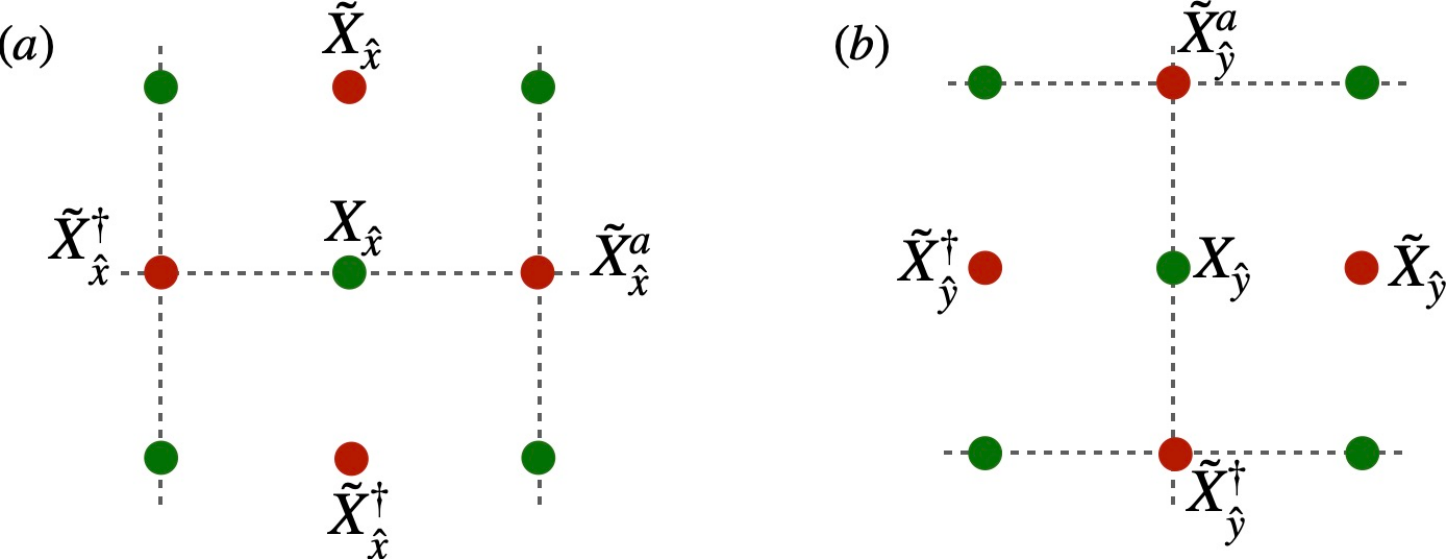}
    \caption{The original toric-code spins live on links (green), while the new gauge potentials live on vertices and plaquette centers (red). Each red site hosts two qubits, denoted $X_{\hat{x}}$ and $X_{\hat{y}}$. (a) The Guass law after gauging $S_{\hat{x}}$. (b) The Guass law after gauging $S_{\hat{y}}$.}
    \label{fig:gauge1}
\end{figure}

Now we promote $S_{\hat{x}}$ and $S_{\hat{y}}$ to gauge symmetries following Ref.~\cite{williamson19}. After introducing dynamical gauge fields, only gauge-invariant operators are physical. Concretely, we add auxiliary $\Z_N$ spins that serve as gauge fields. Since we have two set of symmetries $S_{\hat{x}},S_{\hat{y}}$, we introduce two sets of the gauge fields. Only the spins on ${r,\hat{x}}$ links are acted upon by $S_{\hat{x}}$ (and analogously, spins on ${r,\hat{y}}$ links by $S_{\hat{y}}$); these ${r,\hat{i}}$ links form a square lattice. The new \textit{gauge} spins live on the bonds of this square lattice, which are naturally indexed by the sites and plaquette centers of the original square lattice illustrated in Fig.~\ref{fig:gauge1}. We denote them by $\tilde X/\tilde Z_{r,\hat{i}}$ and $\tilde X/\tilde Z_{p,\hat{i}}$, where $r$ labels original-lattice vertex, $p$ labels original-lattice plaquette centers, and $\hat{i}=\hat{x},\hat{y}$ indicates the gauge potential/electric field that minimally couples to (gauges) the corresponding $S_{\hat{x}}$ or $S_{\hat{y}}$ symmetry charge in Eq.~\eqref{eqn:SinZN}. We label a plaquette by the coordinates of its center. The lattice geometry is illustrated in Fig.~\ref{fig:gauge1}.

As the original spins serve as gauge-field sources, we enforce Gauss’s-law constraints:
 \begin{equation}
	 \begin{gathered}		 X_{r,\hat{x}}\tilde{X}_{r,\hat{x}}^\dag\tilde{X}^{a}_{r+e_x,\hat{x}}\tilde{X}_{r+\frac{e_x}{2}-\frac{e_y}{2},\hat{x}}^\dag\tilde{X}_{r+\frac{e_x}{2}+\frac{e_y}{2}, \hat{x}}=1\\ X_{r,\hat{y}}\tilde{X}_{r,\hat{y}}^\dag\tilde{X}^{a}_{r+e_y,\hat{y}}\tilde{X}_{r-\frac{e_x}{2}+\frac{e_y}{2},\hat{y}}^\dag\tilde{X}_{r+\frac{e_y}{2}+\frac{e_x}{2}, \hat{y}}=1
	 \end{gathered}
	 \label{eq:gat}
 \end{equation}
 for all $r$.
 
The plaquette term is similarly updated by minimal coupling:
\begin{equation}
	\tilde{B}_p= B_r\tilde{Z}_{p,\hat{x}}\tilde{Z}_{p,\hat{y}}^\dag
	\label{}
\end{equation}

The final step is to add plaquette interactions for the new gauge fields:
\begin{align}
    -K'\sum_r \tilde{Z}_{r,\hat{x}}\tilde{Z}_{r+\frac{e_x}{2}+\frac{e_y}{2},\hat{x}}\tilde{Z}_{r+\hat{y},\hat{x}}^\dag \tilde{Z}_{r-\frac{e_x}{2}+\frac{e_y}{2},\hat{x}}^{a\dag} +\text{h.c.}\nonumber\\
       -K'\sum_r \tilde{Z}^a_{r,\hat{y}}\tilde{Z}_{r+\frac{e_x}{2}+\frac{e_y}{2},\hat{y}}\tilde{Z}_{r+e_y,\hat{y}}^\dag \tilde{Z}_{r-\frac{e_x}{2}+\frac{e_y}{2},\hat{y}}^{\dag} +\text{h.c.}.
	\label{}
\end{align}
	\begin{figure}[h]
    \centering
   \includegraphics[width=0.35\textwidth]{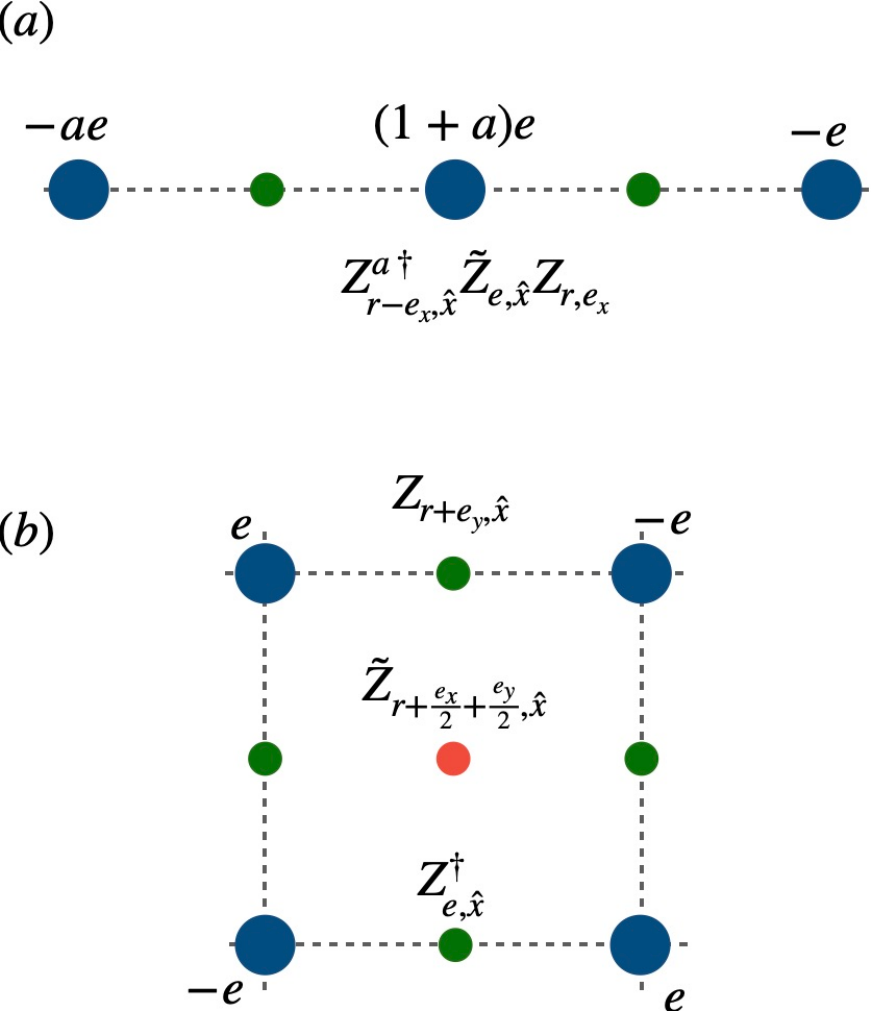}
    \caption{(a) The charge excitation created by the elementary building block $Z_{r-e_x,\hat{x}}^{a\dag},\tilde{Z}_{r,\hat{x}},Z_{r,\hat{x}}$.
(b) The charge excitation created by the elementary building block $Z_{r+e_y,\hat{x}},\tilde{Z}_{r+\frac{e_x}{2}+\frac{e_y}{2},\hat{x}}Z_{r,\hat{x}}^\dag$.
Here, blue dots label the positions of the excitations, with the charge units indicated.}
    \label{fig:gauge2}
\end{figure}

To analyze excitation mobility, we construct gauge-invariant string operators that satisfy the constraint in Eq.~\eqref{eq:gat}. For charge e-strings built from $Z$ or $\tilde{Z}$, Gauss’s law requires specific combinations to maintain gauge invariance. The elementary building blocks are illustrated as Fig.~\ref{fig:gauge2}:
\begin{equation}
Z_{r-e_x,\hat{x}}^{a\dag} 	\tilde{Z}_{r,\hat{x}} Z_{r,\hat{x}}, ~~\tilde{Z}_{r,\hat{y}}Z_{r-e_y,\hat{y}}^{a\dag} Z_{r,\hat{y}}
	\label{eqn:Zsite}
\end{equation}
and
\begin{equation}
	Z_{r+{e_y},\hat{x}} \tilde{Z}_{r+\frac{e_x}{2}+\frac{e_y}{2},\hat{x}} Z_{r,\hat{x}}^\dag,\quad \tilde{Z}_{r+\frac{e_y}{2}+\frac{e_y}{2},\hat{y}}Z_{r+e_x,\hat{y}} Z_{r,\hat{y}}^\dag
	\label{eqn:Zplaq}
\end{equation}
These operators do not commute with nearby $A_r$ and create $e$ excitations in the pattern illustrated in Fig.~\ref{fig:gauge2}. These building blocks produce charge patterns that obey both ordinary and exponential charge conservation, as we will demonstrate shortly.

	\begin{figure}[h]
    \centering
   \includegraphics[width=0.45\textwidth]{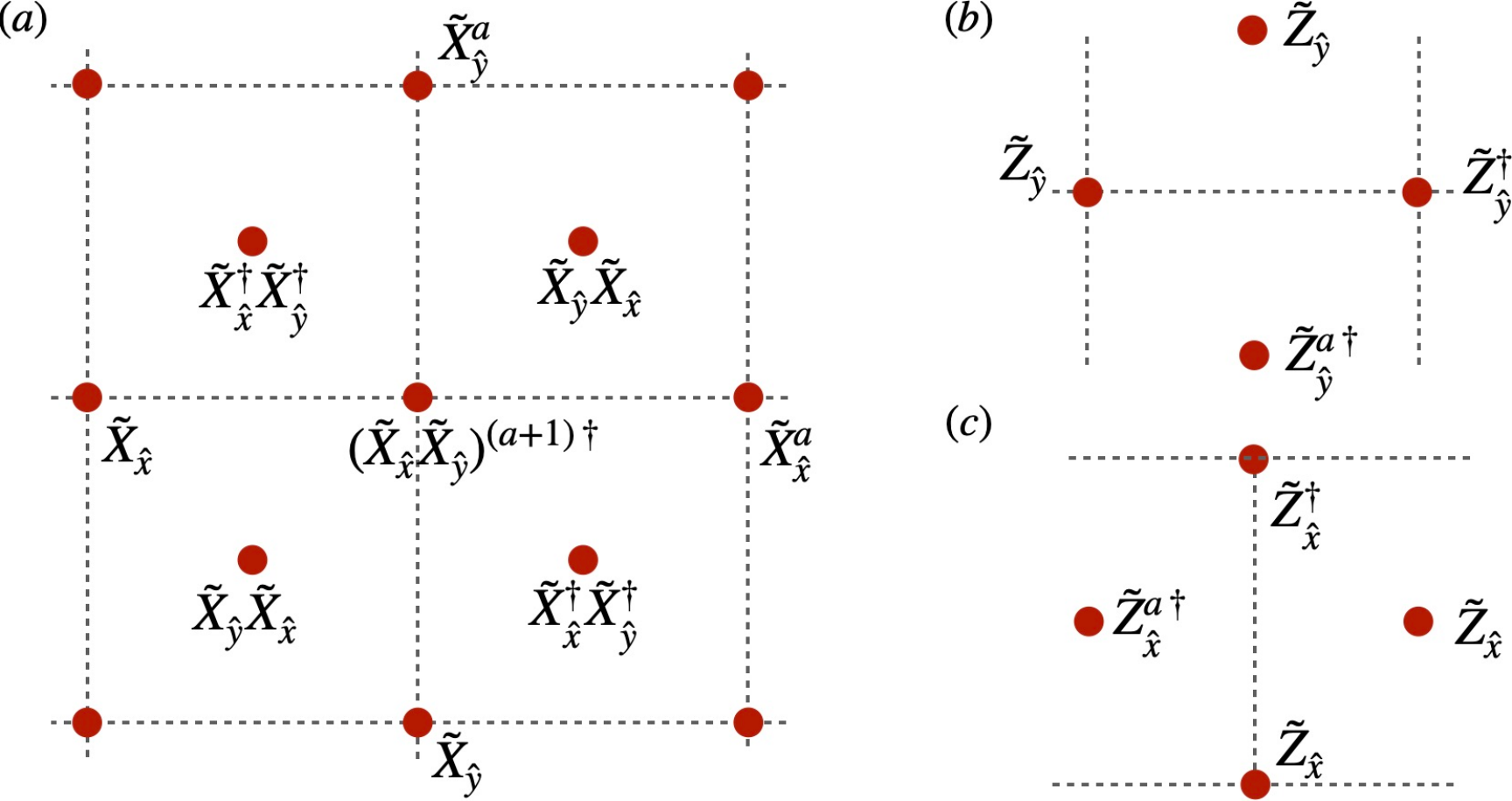}
    \caption{New stabilizer code after symmetry gauging: (a) charge stabilizer $q$; (b–c) flux stabilizers $B^x$ and $B^y$.}
    \label{fig:gauge3}
\end{figure}

We now consider the gauged theory obtained by gauging the exponential symmetry of the toric code in Eq.~\eqref{eqn:SinZN}, thereby imposing the Hilbert–space constraint in Eq.~\eqref{eq:gat}. Under this gauging, the original $X$ qubits on the $x$–$y$ links are traded for $\tilde X$ variables living on vertices and plaquettes. In this representation, the charge stabilizer $A_\vr$ takes the form shown in Fig.~\ref{fig:gauge3}(a), while the new gauge-flux operator is given in Fig.~\ref{fig:gauge3}(b)-(c). Adopting the conventions of Sec.~\ref{sec:two}, these stabilizers naturally organize into a scalar-charge theory upon the following redefinitions:
\begin{eqnarray}
e^{iE^x_{x}}=X_{{r},\hat{x}}, ~e^{iE^x_{y}}=X_{{r}+\frac{{e_x}}{2}+\frac{e_y}{2},\hat{x}},\nonumber\\
e^{iE^y_{x}}=X_{{r}+\frac{e_x}{2}+\frac{e_y}{2},\hat{y}}, ~e^{iE^y_{y}}=X_{{r},\hat{y}},
\end{eqnarray}
Here ${r}$ labels the vertex sites. The generator of gauge transformations is the generalized Gauss law:
\begin{eqnarray}
    D_1 E^x_{x}+D_2 E^y_{y} +D_0(E^y_{x}+E^x_{y})= q,
\end{eqnarray}
Here $q$ denotes the scalar charge.
We define the `lattice derivative' operators as follows:
\begin{eqnarray}
    &D_1f_r=-f_{r+\hat{x}}-a f_{r-\hat{x}}+(a+1)f_r,\nonumber\\
&D_2f_r=-f_{r+\hat{y}}-a f_{r-\hat{y}}+(a+1)f_r,\nonumber\\
&D_0 f_r=f_{r+\hat{x}}-f_r - (f_{r+\hat{y}+\hat{x}}-f_{r+\hat{y}})
\end{eqnarray}
Define the gauge-invariant magnetic flux operator as:
\begin{align} 
&B^x=d_1 A^x_{y} - \nabla_y A^x_{x},~B^y=\nabla_x A^y_{y} - d_2 A^y_{x},\nonumber\\
& d_1 f_r=f_{r}-a f_{r-\hat{x}},\nonumber\\
&d_2 f_r=f_{r}-a f_{r-\hat{y}},
\end{align}

After some algebra, one finds that the charge and magnetic-flux operators obey the following conservation laws on an $L_x \times L_y$ periodic lattice:
\begin{align}
&\sum_{{r}}  a^{L_x-x} B^x_{{r}} = 0,\quad
\sum_{{r}} a^{L_y-y} B^y_{{r}} = 0 ,\nonumber\\
&\sum_{{r}} q_{{r}}=0 \quad
\sum_{{r}} a^{x}q_{{r}}=0, ~ \sum_{{r}} a^{y} q_{{r}}=0.
\end{align} 
In summary, this theory also exhibits exponential charge conservation. In the original toric code, the $e$ charge is globally conserved; here, however, the string operators that move $e$ are themselves charged under $S_{\hat{x}}$ and $S_{\hat{y}}$, so the construction is naturally a symmetry-enriched toric code. Upon further gauging $S_{\hat{x}}$ and $S_{\hat{y}}$, the charge excitations acquire new mobility constraints, and the resulting gauge theory illustrated in Fig.~\ref{fig:gauge3} exhibits both global and exponential charge conservation.
Further variations of the exponential gauge theory can be obtained via anyon condensation. In particular, by following the anyon-condensation web of Ref.~\cite{delfino2024anyon} and imposing $E_x^y=0$ and $E_y^x=0$, one recovers the stabilizer code described in Sec.~\ref{sec:two}.

\bibliography{ref.bib}

\end{document}